\documentclass[iop,revtex4]{emulateapj}
\usepackage{lineno}

\usepackage{lscape}

\begin{document}

\renewcommand{\topfraction}{1.0}
\renewcommand{\bottomfraction}{1.0}
\renewcommand{\textfraction}{0.0}

\shorttitle{Speckle interferometry at SOAR}
\shortauthors{Tokovinin et al.}

\title{Speckle interferometry at SOAR in 2018 }

\author{Andrei Tokovinin}
\affil{Cerro Tololo Inter-American Observatory, Casilla 603, La Serena, Chile}
\email{atokovinin@ctio.noao.edu}

\author{Brian D. Mason}
\affil{U.S. Naval Observatory, 3450 Massachusetts Ave., Washington, DC, USA}
\email{brian.d.mason@navy.mil}
\author{Rene A. Mendez}
\affil{Universidad de Chile,  Casilla 36-D, Santiago, Chile}
\email{rmendez@u.uchile.cl}
\author{Elliott P. Horch\footnote{Adjunct Astronomer, Lowell Observatory} }
\affil{Department of Physics, Southern Connecticut State University, 501 Crescent Street, New Haven, CT 06515, USA}
\email{horche2@southernct.edu}
\author{Cesar Brice\~no}
\affil{Cerro Tololo Inter-American Observatory, Casilla 603, La Serena, Chile}
\email{cbriceno@ctio.noao.edu}

\begin{abstract}
The  results of  speckle  interferometric observations  at  the 4.1  m
Southern Astrophysical  Research Telescope  (SOAR) in 2018  are given,
totaling  3097 measurements  of 2427  resolved pairs  with separations
from  11  mas  to  5\farcs9 (median  0\farcs15,  magnitude  difference
up to  7  mag)  and  non-resolutions  of 624  targets.   This  work
continues our long-term  speckle program. Its main goal  is to monitor
orbital  motion of  close  binaries, including  members of  high-order
hierarchies  and  {\it Hipparcos}  pairs  in  the solar  neighborhood.
Also,  pre-main-sequence  stars  in  the Orion  OB1  association  were
surveyed, resolving  26 out  of 118 targets.   In addition,  we report
discovery of 35  new companions among field visual  multiples (some of
which  are likely optical)  and first-time  resolutions of  another 31
pairs.  By  combining the measurements  given here with  the published
ones, we  computed 76  orbits for the  first time and  updated orbital
elements of 34 visual binaries.  Their periods range from 0.65 to 1100
years,  and their  quality varies  from first  tentative  solutions of
grade 5 to accurate elements of grades 1 and 2.  Finally, a list of 53
spurious pairs discovered by various techniques and unresolved at SOAR
is given.
\end{abstract} 

\section{Introduction}
\label{sec:intro}

We report  here a  large set of  double-star measurements made  at the
4.1 m  Southern  Astrophysical  Research  Telescope  (SOAR)  with  the
speckle camera,  HRCam.  This paper continues the  series published by
\citet[][hereafter   TMH10]{TMH10},  \citet{SAM09},  \citet{Hrt2012a},
\citet{Tok2012a},  \citet{TMH14},  \citet{TMH15},
\citet{SAM15}, and \citet{SAM17}.  Most data were taken during 2018,
but some older, unpublished measurements are presented here as well.

Section~\ref{sec:obs} reviews all speckle programs executed at SOAR in
2018, recent  changes to the observing procedure,  and the astrometric
calibration. The results are presented in Section~\ref{sec:res} in the
form of electronic tables archived by the journal. We also discuss new
resolutions,  provide  a  large  list  of new  orbital  elements,  and
indicate    likely    spurious    pairs.    A   short    summary    in
Section~\ref{sec:sum} closes the paper.

\section{Observations}
\label{sec:obs}

\subsection{Observing programs}

During  2018, HRCam  (see  Sect.~\ref{sec:inst}) was  used to  execute
several  observing programs, some  with common  (overlapping) objects.
Table~\ref{tab:programs}  gives  an  overview  of these  programs  and
indicates which observations are  published in the present paper. Here
is a brief description of these programs.

{\it  Orbits} of resolved  binaries are  of fundamental  importance in
various  areas of astronomy,  e.g. for  direct measurement  of stellar
masses, binary statistics, astrometry, and objects of special interest
such as binaries hosting exo-planets. Observations of tight pairs with
fast motion, mostly nearby dwarfs, are prioritized at SOAR.  Recently,
\citet{Mason2018}  published orbits of  low-mass red  dwarfs partially
based  on our data.  However, ``classical''  visual binaries  are also
observed {\bf with appropriate temporal sampling} to  improve their orbits. The Sixth Catalog of
Visual  Binary Star  Orbits,  VB6 \citep{VB6},  contains a  substantial
fraction of  poorly determined,  low-grade orbits based  on inaccurate
and/or sparse  visual micrometric  measures. This situation  is slowly
improving.  Our  work added  202 orbits to  VB6, published  between 2017
and 2018. More orbits are given here in Section~\ref{sec:orbits}.

{\it  Hierarchical  systems}  of   stars  challenge  the  theories  of
binary-star formation.  Better  observational data on their statistics
and   architecture   (orbits,   relative  inclinations)   are   needed
\citep{MSC}.   Many hierarchies  have  been discovered  at SOAR  using
HRCam,  and we  are following  their orbital  motion.   An interesting
class of double twins --- triple systems  with quasi-coplanar orbits and
moderate    period    ratios ---    has    been    recently    identified
\citep{twins}.  This paper adds  several newly  discovered hierarchies
and several orbits of  subsystems.

{\it Hipparcos binaries} within 200\,pc  are monitored with the aim of
determining orbits and  masses for stars in a  wide range of effective
temperatures      and      metallicities,      as     outlined      by
\citet{Horch2015,Horch2017,Horch2019}.   The  southern  part  of  this
sample is addressed at SOAR \citep{Mendez2017}.  This program overlaps
with  the  general work  on  orbits.   Accurate  parallaxes of  visual
binaries,  soon   to  be  measured   by  {\it  Gaia},   combined  with
good-quality  orbits,  will  allow  accurate measurements  of  stellar
masses. However,  it is naive to  expect that {\it  Gaia} will deliver
precise parallaxes without knowledge of the orbits, as parallactic and
orbital motions are coupled.  The current {\it Gaia} data release, DR2
\citep{Gaia}, contains  examples of biased parallaxes  of close visual
pairs   owing  to  this   coupling.\footnote{For  example,   the  {\it
    Hipparcos} parallax of HIP~4869, a  visual binary with a period of
  28 years,  is 47.3$\pm$1.2 mas,  matching the dynamical  parallax of
  44.5 mas  and the  {\it Gaia} parallax  of its common  proper motion
  companion  NLTT 3509,  46.9$\pm$0.1 mas.   Yet, the  {\it  Gaia} DR2
  parallax of 65.8$\pm$0.6 mas is obviously biased, while its large error
  indicates the inadequacy of the current 5-parameter astrometric
  model that does not account for the orbit.}

{\it  Binarity in  the  Orion  OB1 association}  was  studied in  2016
January  (PI C.B.) using  the new  catalog of  pre-main-sequence (PMS)
stars published by \citet{CIDA}.   Statistical analysis of this survey
will  be  presented in  a  forthcoming  paper.   Here we  provide  the
observational  data, namely  new close  binaries  and non-resolutions.
Owing to the faintness of these targets, the laser guide star (see the
instrument  description in Sect.~\ref{sec:inst})  was used  to sharpen
the  images  and thus  increase  the  sensitivity  at the  expense  of
efficiency.   However,  the new  CCD  used  in  HRCam since  2017  has
improved the magnitude limit to the point where several of these stars
could be  re-observed and confirmed  without the help of  laser, under
good seeing.

{\it Kepler  multi-periodic stars}  in the Upper  Scorpius association
were assumed  to be  close binaries. Indeed,  we were able  to resolve
most of  them and published  our results in \citet{Sco-M}.   This work
raised our awareness of the  poor census of binaries in this important
young stellar  aggregate.  We  continue to survey  a large  and nearly
complete sample  of PMS stars  in this group  and hope to  publish the
results soon.

\begin{deluxetable}{ l l l l  } 
\tabletypesize{\scriptsize}    
\tablecaption{Observing programs executed with HRCam in 2018
\label{tab:programs} }                    
\tablewidth{0pt}     
\tablehead{ \colhead{Program}  &
\colhead{PI}  &  
\colhead{$N$} & 
\colhead{Publ.\tablenotemark{a}} 
}
\startdata
Orbits                & Mason, Tokovinin    & 1130 & Yes \\
Hierarchical systems  & Tokovinin           & 258 & Yes \\
Hipparcos binaries    & Mendez, Horch       & 648 & Yes \\
Binaries in Ori OB1   & Brice\~no           & 155 & Yes \\
Kepler multi-periodic & Tokovinin, Brice\~no & 129 & Pub \\
Neglected binaries    & R.~Gould, Tokovinin & 863  & Yes \\
Young associations    & Brice\~no, Tokovinin & 227 & No \\
Nearby K,M dwarfs     & J. Winters, D. Nusdeo & 100 & No \\
Eclipsing binaries    & D. Martin            & 34 & No \\
TESS follow-up        & C. Ziegler           & 90  & No \\
Young moving groups   & A. Mann              & 345  & No \\ 
Stars with RV trends  & B. Pantoja           & 39  & No
\enddata
\tablenotetext{a}{This columns indicates whether the results are
  published here (Yes), previously (Pub), or deferred to future
  papers (No). }
\end{deluxetable}

{\it Neglected  binaries} with  small separations from  the Washington
Double Star Catalog, WDS \citep{WDS} are observed with a low priority,
as a ``filler''.  Lists of pairs in need of fresh data are provided by
R.~Gould  {\bf (private  communication,  2018).} A  fraction of  these
stars are interesting because they  are presently very tight, near the
periastron of their orbits.  Some of these pairs turned out to contain
additional  previously  unknown  components.   Owing to  the  improved
observing efficiency of HRCam, the regular program in 2018 March-April
used only  part of the  allocated time.  Two ``filler''  programs were
improvised, namely measurements of southern binaries from the WDS with
separations between 0\farcs1 and  0\farcs4 that were never observed at
SOAR  and observations  of  wide  physical pairs  in  search of  close
subsystems.   These  programs led  to  the  discovery  of several  new
hierarchical systems  and helped to  pinpoint a number of  false pairs
that pollute the WDS catalog.

{\it Nearby  K and M  dwarfs} were observed  on request from  T. Henry
(PIs J.  Winters and  D.  Nusdeo).  A number of  binaries were  resolved,
apparently for the first time.

Several programs initiated in 2018  are still in progress, such as the
high-resolution follow-up of TESS objects of interest, survey of stars
in young moving groups, and search for tertiary companions to low-mass
eclipsing binaries.

\subsection{Instrument and observing procedure}
\label{sec:inst}

The   observations  reported   here  were   obtained  with   the  {\it
  high-resolution camera} (HRCam) -- a fast imager designed to work at
the 4.1 m SOAR telescope \citep{HRCAM}.  The camera was mounted on the
SOAR  Adaptive Module \citep[SAM,][]{SAM}.   However, the  laser guide
star  of SAM was  not used  (except in  2016 January);  the deformable
mirror   of  SAM   was  passively   flattened  and   the   images  are
seeing-limited.   In most observing  runs, the  median image  size was
$\sim$0\farcs6.   The SAM module  contains the  atmospheric dispersion
corrector (ADC).   The transmission curves of HRCam  filters are given
in             the             instrument            manual.\footnote{
  \url{http://www.ctio.noao.edu/soar/sites/default/files/SAM/\-archive/hrcaminst.pdf}}
We  used  mostly  the  Str\"omgren  $y$ filter  (543/22\,nm)  and  the
near-infrared $I$  filter (824/170\,nm). A  few measures were  made in
the $V$ (517/84\,nm) and $R$ (596/121\,nm) filters. 
{\bf The detector is the electron multiplication CCD iXon-888.}
Observations in 2016 used a different detector (Luca-DL), and the $I$-band response was 788/132\,nm. 

For each observing  run, a unified observing list  of objects from all
programs was  prepared.  It  contains accurate coordinates  and proper
motions  (PMs) to  allow for  precise pointing  of the  telescope. The
slews  are commanded  from the  custom  observing tool  that helps  to
maximize the observing  efficiency. When the slew angle  is small, the
next object  is acquired  almost immediately.  Most  observations were
taken  in the  narrow  3\arcsec ~field  with  the 200$\times$200  {\bf
  pixels}  region  of interest  (ROI),  without  binning,  in the  $I$
filter;  the $y$  filter was  used mostly  for brighter  and/or closer
pairs.   The pixel  scale is  0\farcs01575  and the  exposure time  is
normally 24\,ms  (it is limited  by the camera readout  speed).  Pairs
wider  than  $\sim$1\farcs4  are  observed in  a  400$\times$400  {\bf
  pixels} ROI,  and the widest  pairs are sometimes recorded  with the
full field of 1024 pixels (16\arcsec) and the 2$\times$2 binning.  The
binning is used mostly for the  fainter targets; it does not result in
the loss  of resolution in  the $I$ band,  which ranges from 40  to 45
mas, depending on  the magnitude and conditions.  Bright  stars can be
resolved and  measured below the formal diffraction  limit (an example
is  given  below   in  Sect.~\ref{sec:orbits}).   The  resolution  and
contrast  limits  of HRCam  are  further  discussed  in TMH10  and  in
the previous papers of this series.

\begin{figure}
\epsscale{1.1}
\plotone{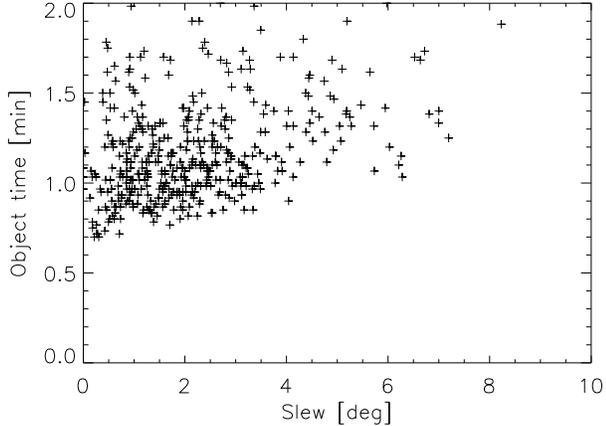}
\caption{Correlation between  observing time  per object and  the slew
  distance for  the night of 2018  April 3/4. The  vertical axis plots
  time between observations of successive objects.
\label{fig:slew} }
\end{figure}

On the night  of 2018 April 3/4, a total of  466 targets have been observed
during  10.6  hours.  The   average  time  between  targets  was  1.36
min.  Figure~\ref{fig:slew} illustrates  the   correlation between
the  target time and  the slew  distance; larger  slews take  a longer
time. Typically, HRCam covers about 300 targets in one night.

In 2018, we implemented the automatic selection of reference stars for
measuring the  speckle transfer function. Their general  list is based
on the {\it Hipparcos} catalog \citep{HIP}, with Hp magnitudes between
5 and 7 and excluding  known binaries.  For each target, the observing
tool  offers five  closest references  from this  list and  points the
telescope to the selected reference,  if asked.  In this way, there is
no need to include reference  stars in the observing program, and they
can be chosen flexibly.   Binaries with magnitude difference $\Delta m
> 1$  mag  and unresolved  targets  (e.g.   from  the binarity  survey
program) are  used as reference during  data processing \citep[see][]{SAM15},
so   special  observations   of  reference   stars  are   needed  only
occasionally.

\begin{figure}
\epsscale{1.1}
\plotone{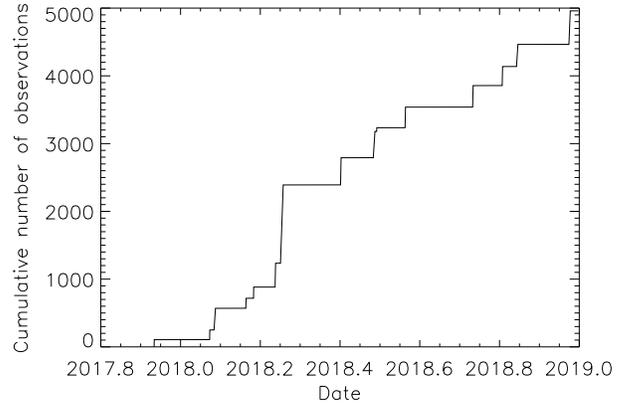}
\caption{Cumulative plot  of the number of HRcam  observations at SOAR
  during 2018 (all programs).
\label{fig:cum} }
\end{figure}

The first  observations reported here were obtained  in 2017 December,
and  the last  in  2018  December.  HRCam  was  used during  scheduled
observing runs, but also in parts of engineering nights available from
other  work.   Figure~\ref{fig:cum}  plots  the cumulative  number  of
observations executed  during this year, which reaches  almost 5000. The
largest number of objects was  covered during four scheduled nights in
March--April 2018.

\subsection{Data processing and calibration}
\label{sec:dat}

\begin{figure}[ht]
\epsscale{1.0}
\plotone{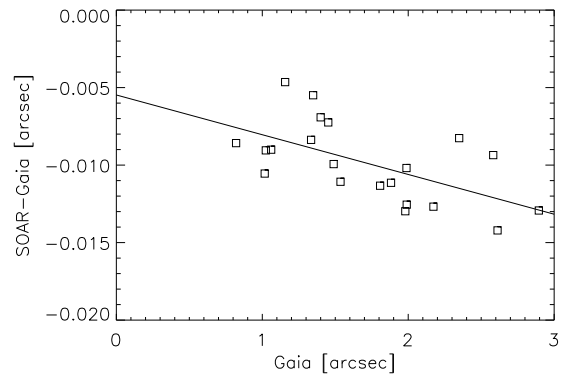}
\caption{Comparison  between  SOAR   and  {\it  Gaia}  separations  of
  calibration    binaries.    The    line    is    a    linear    fit given by
  Equation~\ref{eq:rho}.
\label{fig:gaia} }
\end{figure}

The data processing  is described in TMH10 and  \citet{HRCAM}.  We use
the standard speckle interferometry technique based on the calculation
of the power spectrum  and the speckle auto-correlation function (ACF)
derived from  it.  Companions are  detected as secondary peaks  in the
ACF and/or as fringes in the power spectrum.  Parameters of the binary
and  triple stars  (separation  $\rho$, position  angle $\theta$,  and
magnitude  difference  $\Delta  m$)  are determined  by  modeling  the
observed  power spectrum.   Additionally, the  true quadrant  is found
from the shift-and-add images, whenever possible. 

The pixel scale  and angular offset are determined  by observations of
several relatively wide calibration  binaries. Their motion is modeled
based  on previous  observations at  SOAR, with  individual  scale and
orientation  corrections  for  each  observing  run.  The  models  are
adjusted iteratively. The latest adjustment of 65 calibrators was done
in 2017  November. Typical rms  deviations of observations  from these
models are 0\fdg2 in angle and 1 to 3 mas in separation.

The adopted calibration procedure assures good internal consistency of
the SOAR  speckle astrometry  but does not  preclude the  existence of
global systematic  errors. We compared  a subset of 21  calibrators to
the  {\it  Gaia} astrometry  provided  in  the  DR2 \citep{Gaia}.  The
separations   range  from   0\farcs82  to   2\farcs2   (the  remaining
calibrators are not resolved in  the DR2).  We computed the {\it Gaia}
position  angles  and  separations  for J2000  from  the  coordinates,
corrected them  for precession  in angle to  the epoch of  2015.5, and
compared to the positions predicted by our models for the same date.

The comparison reveals a  small, but measurable difference between the
SOAR  and {\it  Gaia}  ``systems''. The  position angles  $\theta_{\rm
  SOAR}$  have, on  average,  an  offset of  $-$0\fdg17,  with an  rms
scatter of  0\fdg12 around  this value (or  3.1 mas in  the tangential
direction).   The   scatter  decreases  with   separation.   The  SOAR
separations  are smaller  compared to  those  from {\it  Gaia}, and  a
linear trend is found:
\begin{equation}
\rho_{\rm SOAR} - \rho_{Gaia} \approx -0.0054 - 0.0025 \rho_{Gaia} ,
\label{eq:rho}
\end{equation}
as shown  in Figure~\ref{fig:gaia}.  The rms scatter  around this line
is 2.05\,mas. 

The  systematical errors  of the  HRCam astrometry  are less  than the
declared calibration accuracy, 0.5\% in  scale and 0\fdg5 in angle. We
do not apply  these corrections to the data  presented here but rather
prefer  to  keep  the  HRCam   astrometry  on  the  same  system,  for
consistency.  When  the {\it  Gaia} DR4 containing  a large  volume of
double-star astrometry  becomes available,  we will repeat  and extend
its comparison  with HRCam and  will determine the  final corrections.
At  present,   it  cannot   be  excluded  that   the  trend   seen  in
Figure~\ref{fig:gaia} is not caused,  at least partially, by errors in
the {\it Gaia} data.  The optics  of HRcam has a cubic distortion that
reduces the  pixel scale off-axis.   However, this distortion  is very
small: the  relative pixel scale is  reduced only by  $3 \times 10^{-5}$ for a
 4\arcsec ~offset.

\section{Results}
\label{sec:res}

\subsection{Data tables}

\begin{figure}
\epsscale{1.1}
\plotone{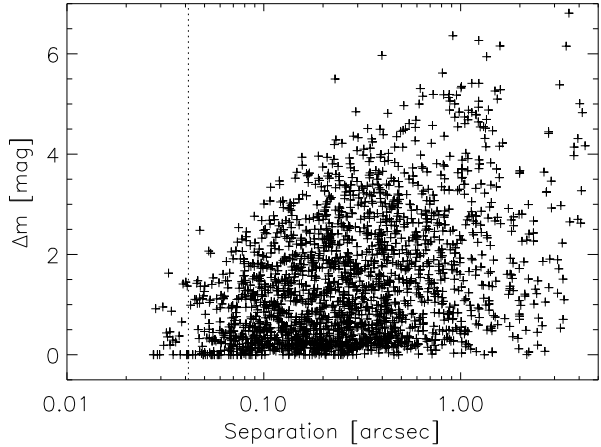}
\caption{Magnitude difference in the $I$ band vs. separation for pairs
  resolved in this filter.  The  vertical dotted line marks the formal
  diffraction limit of 41\,mas.
\label{fig:dm} }
\end{figure}

The results (measures of resolved pairs and non-resolutions) are
presented in almost the same format  as in the previous papers of this
series. The long tables are published electronically; here we describe
their content. To illustrate the resolution and dynamic range of this
data set, we plot in Figure~\ref{fig:dm} magnitude difference
vs. separation for  pairs resolved in the $I$ filter. 

\begin{deluxetable}{ l l  l l }
\tabletypesize{\scriptsize}
\tablewidth{0pt}
\tablecaption{Measurements of double stars at SOAR 
\label{tab:measures}}
\tablehead{
\colhead{Col.} &
\colhead{Label} &
\colhead{Format} &
\colhead{Description, units} 
}
\startdata
1 & WDS    & A10 & WDS code (J2000)  \\
2 & Disc.  & A16 & Discoverer code  \\
3 & Other  & A12 & Alternative name \\
4 & RA     & F8.4 & R.A. J2000 (deg) \\
5 & Dec    & F8.4 & Declination J2000 (deg) \\
6 & Epoch  & F9.4 & Julian year  (yr) \\
7 & Filt.  & A2 & Filter \\
8 & $N$    & I2 & Number of averaged cubes \\
9 & $\theta$ & F8.1 & Position angle (deg) \\
10 & $\rho \sigma_\theta$ & F5.1 & Tangential error (mas) \\
11 & $\rho$ & F8.4 & Separation (arcsec) \\
12 &  $\sigma_\rho$ & F5.1 & Radial error (mas) \\
13 &  $\Delta m$ & F7.1 & Magnitude difference (mag) \\
14 & Flag & A1 & Flag of magnitude difference\tablenotemark{a} \\
15 & (O$-$C)$_\theta$ & F8.1 & Residual in angle (deg) \\
16 & (O$-$C)$_\rho$ & F8.3 & Residual in separation (arcsec) \\
17  & Ref. & A8   & Orbit reference\tablenotemark{b} 
\enddata
\tablenotetext{a}{Flags: 
q -- the quadrant is determined; 
* -- $\Delta m$ and quadrant from average image; 
: -- noisy data. }
\tablenotetext{b}{References to VB6 are provided at
  \url{http://ad.usno.navy.mil/wds/orb6/wdsref.txt}; Tab.7 refers to Table~7 of this paper. }
\end{deluxetable}

Table~\ref{tab:measures}  lists 3097 measures  of 2427  resolved pairs
and  subsystems,  including  the   new  discoveries.   The  pairs  are
identified by  their WDS codes and discoverer  designations adopted in
the WDS catalog \citep{WDS}, as well as by alternative names in column
(3), mostly  from the {\it Hipparcos}  catalog. Equatorial coordinates
for the  epoch J2000 in  degrees are given  in columns (4) and  (5) to
facilitate matching with other catalogs and databases.  In the case of
multiple systems, the position  measurements and their errors (columns
9--12) and  magnitude differences (column 13) refer  to the individual
pairings between  components, not to  their photo-centers.  As  in the
previous papers  of this series,  we list the internal  errors derived
from  the power  spectrum model  and from  the difference  between the
measures obtained from two data cubes. The median {\bf internal} error
is 0.3\,mas, and 95\% of {\bf these} errors  are less than 3\,mas. The real {\bf external)}  errors
are usually  larger, especially  for difficult pairs  with substantial
$\Delta  m$  and/or with  small  separations.   Residuals from  orbits
(Section~\ref{sec:orbits})  and   from  the  models   of  calibrators,
typically between 1 and 5 mas rms, characterize the external errors of
the HRcam astrometry.

The  flags in column  (14) indicate  cases when  the true  quadrant is
determined (otherwise the position angle is measured modulo 180\degr),
when the  photometry of wide  pairs is derived from  the long-exposure
images (this  reduces the bias  caused by speckle  anisoplanatism) and
when the  data are noisy or  the resolutions are  tentative.  {\bf The
  exact definition of noisy data, related to the signal to noise ratio
  in the power  spectrum, is given in TMH10;  such observations have a
  lower  resolution   limit  and  precision.  For   pairs  wider  than
  $\sim$1\arcsec, our estimates  of $\Delta m$ may be  too large owing
  to anisoplanatism and potential  truncation of the companion's image
  in  the narrow  3\arcsec  ~field.   } For  binary  stars with  known
orbits,  the residuals  to  the  latest orbit  and  its reference  are
provided in columns (15)--(17). {\bf The orbits computed in this paper
  are referenced as ``Tab.7''. }

Non-resolutions are reported in Table~\ref{tab:single}. Its first
columns (1) to (8) have the same meaning and format as in
Table~\ref{tab:measures}. Column (9) gives the minimum resolvable
separation when pairs with $\Delta m < 1$ mag are detectable. It is
computed from the maximum spatial frequency of the useful signal in
the power spectrum and is normally close to the formal diffraction
limit $\lambda/D$. The following columns (10) and (11) provide the
indicative dynamic range, i.e. the maximum magnitude difference at
separations of 0\farcs15 and 1\arcsec, respectively. 

\begin{deluxetable}{ l l  l l }
\tabletypesize{\scriptsize}
\tablewidth{0pt}
\tablecaption{Unresolved stars 
\label{tab:single}}
\tablehead{
\colhead{Col.} &
\colhead{Label} &
\colhead{Format} &
\colhead{Description, units} 
}
\startdata
1 & WDS    & A10 & WDS code (J2000)  \\
2 & Disc.  & A16 & Discoverer code  \\
3 & Other  & A12 & Alternative name \\
4 & RA     & F8.4 & R.A. J2000 (deg) \\
5 & Dec    & F8.4 & Declination J2000 (deg) \\
6 & Epoch  & F9.4 & Julian year  (yr) \\
7 & Filt.  & A2 & Filter \\
8 & $N$    & I2 & Number of averaged cubes \\
9 & $\rho_{\rm min}$ & F7.3 & Angular resolution (arcsec)  \\
10 &  $\Delta m$(0.15) & F7.2 & Max. $\Delta m$ at 0\farcs15 (mag) \\
11 &  $\Delta m$(1) & F7.2 & Max. $\Delta m$ at 1\arcsec (mag) 
\enddata
\end{deluxetable}

Table~\ref{tab:measures} contains  about a hundred  pairs resolved for
the  first  time; some  of  those  were  confirmed in    subsequent
observing runs.  Almost as many additional first resolutions belonging
to the  projects led  by other PIs  will be reported  elsewhere (these
pairs are  not published here), while  54 new pairs  in Upper Scorpius
are  published by  \citet{Sco-M}.  In  the following  sub-sections, we
discuss new resolutions in the context of observing programs.

\begin{deluxetable}{ l c  r r r c }
\tabletypesize{\scriptsize}
\tablewidth{0pt}
\tablecaption{New pairs in Orion OB1 
\label{tab:Ori}}
\tablehead{
\colhead{WDS} &
\colhead{CVSO} &
\colhead{$\rho$} &
\colhead{$\theta$} &
\colhead{$\Delta I$} &
\colhead{Conf.}  \\
& & 
\colhead{(arcsec)} & 
\colhead{(deg)} & 
\colhead{(mag)} & 
}
\startdata
05022$-$0408 &267    &  0.312 & 328.6 &   1.3 & \\     
05042$-$0005 &271    &  1.102 & 119.6 &   2.8 & \\     
05067$-$0318 &286    &  1.257 & 251.6 &   4.5 & \\     
05119$-$0157 &324    &  0.675 &  66.1 &   3.6 & * \\     
05204$-$0001 &7      &  0.188 & 205.4 &   1.5 & * \\     
05220$+$0144 &516    &  0.179 & 276.4 &   1.5 & * \\     
05223$+$0201 &530    &  0.295 & 229.9 &   0.6 & * \\     
05245$+$0148 &2001   &  0.640 &  95.8 &   4.1 & \\     
05253$-$0158 &33     &  2.122 &  46.4 &   4.6 & \\     
05257$+$0145 &35     &  1.283 & 272.7 &   2.0 & \\     
05261$-$0209 &840    &  0.187 & 198.2 &   0.6 & * \\     
05277$+$0312 &985    &  0.684 & 285.7 &   2.2 & \\     
05294$+$0136 &1169   &  0.563 & 279.3 &   1.0 & \\     
05296$-$0135 &65     &  1.502 & 152.6 &   0.7 & \\     
05318$-$0155 &98     &  0.165 & 298.5 &   1.8 & * \\     
05319$-$0045 &1328   &  0.572 & 309.0 &   0.7 & \\     
05335$-$0132 &120    &  3.720 & 314.2 &   2.0 & \\     
05342$-$0009 &130    &  1.269 & 106.5 &   1.8 & \\     
05345$-$0204 &1506   &  2.469 &  32.1 &   3.7 & \\     
05352$-$0043 &1577   &  0.185 & 267.7 &   3.4 & * \\     
05353$-$0050 &141    &  0.086 &  68.9 &   0.0 & * \\     
05356$-$0143 &1620   &  0.458 & 135.1 &   2.6 & \\     
05357$+$0021 &1633   &  0.278 & 143.4 &   0.6 & * \\     
05375$-$0048 &155    &  1.981 & 128.6 &   1.1 & \\     
05379$-$0009 &1789   &  1.510 & 225.2 &   1.5 & \\     
05397$-$0035 &173    &  0.161 & 157.4 &   0.3 & *     
\enddata
\end{deluxetable}

\subsection{New pairs in Orion OB1}
\label{sec:Ori}

\begin{figure}
\epsscale{1.1}
\plotone{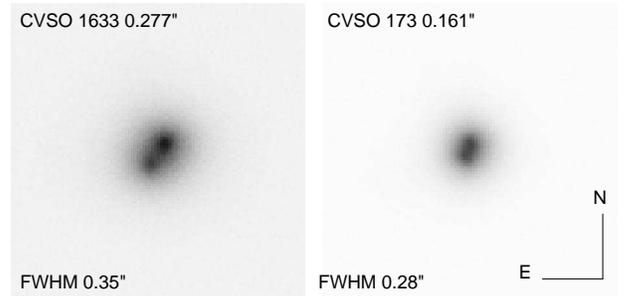}
\caption{Centered images of two  newly resolved close binaries in Orion
  OB1, displayed  on arbitrary negative scale.  The binary separation  and FWHM
  resolution are indicated.
\label{fig:Ori} }
\end{figure}

In 2016 January 16--18, we  surveyed young low-mass stars in the Orion
OB1 association.   We targeted 150  objects amongst the  brightest ($V
\le 15$)  of the  2062 T Tauri  stars (TTS) reported  by \citet{CIDA}.
This  sample includes  74 young  stars in  the $\sim  5$ Myr  old OB1b
sub-association and another 74  in the older OB1a sub-association; the
latter  are  distributed  as  follows:  42  are  part  of  the  widely
distributed ``young field'' population of OB1a ($\sim 11$ Myr), 17 are
members of  the 25  Ori cluster and  8 of  the HD 35762  cluster (both
$\sim 8$  Myr old),  and 7 belong  to the  HR 1833 cluster  ($\sim 13$
Myr).  There  are 30  accreting classical TTS  (CTTS) stars  among the
sample,  111   non-accreting  weak-line  TTS  (WTTS)  and   7  of  the
newly-defined   C/W   class,   objects   with   accretion   properties
intermediate between CTTS  and WTTS, possibly because they  are in the
process of ending their accretion phase.  Roughly half of the CTTS are
located in OB1b. This is by design, in order to have similar number of
accreting  TTS in  both  regions {\bf  for  statistical comparison  of
  multiplicity of accreting and  non-accreting stars.} In reality, the
younger OB1b region  contains roughly twice as many  CTTS as the older
OB1a (which includes the three clusters mentioned above).

As most targets were quite faint, we used the SAM laser guide star for
partial compensation of turbulence to  get sharper images. The AO loop
did not compensate for the  tilts; instead, the individual frames were
centered and co-added in the  data processing.  The good seeing during
these observations and the AO compensation resulted in the median full
width at half maximum (FWHM)  of re-centered images of 0\farcs33 (best
0\farcs25), while  the site monitor  reported seeing from  0\farcs5 to
1\arcsec ~during these observations. In the  morning, when Orion was
too  low,  we  observed  stars  in the  young  association  $\epsilon$
Chamaeleontis \citep{Cha}.

Data  cubes  were  taken  with  HRCam  in  the  $I$  filter  (response
788/132\,nm for the  Luca DL camera used in  2016) with exposure time
of 0.1 or 0.2 s per  frame, longer than usual, and with the 2$\times$2
binning. Data  cubes with a  smaller field and shorter  exposures were
also acquired;  they were useful  for stars brighter than  $I=12$ mag.
The  data  were  processed  by  the  standard  speckle  pipeline.   In
addition,      we     examined     average      re-centered     images
(Figure~\ref{fig:Ori}) where the  smooth component approximated by the
Moffat  function was  subtracted.  This  helped to  detect  or confirm
faint companions at larger separations.   In 2017 and 2018, some newly
discovered  close binaries  were re-measured  without the  laser image
sharpening  because  the HRCam  used  a  new  CCD camera  with  better
sensitivity.

Statistical  analysis  of  the   binary  population  in  the  Ori  OB1
association  is  beyond  the  scope   of  this  paper.   It  will  use
seeing-limited  images  and {\it  Gaia}  astrometry  to address  wider
binaries. Here  we only report the  speckle results. The  PMS stars in
Ori OB1 are identified by  their CVSO numbers \citep{CIDA} in the main
tables. A summary of 26 new  pairs discovered in 2016.04 in Ori OB1 is
given in  Table~\ref{tab:Ori}.  It contains the WDS  code derived from
the J2000 coordinates (naturally, these objects are not yet present in
the  actual  WDS),  CVSO  number,  separation,  angle,  and  magnitude
difference in the $I$ band.   An asterisk in the last column indicates
subsequent confirmation in 2017--2018. Interestingly, the closest pair
CVSO~141  shows  some orbital  motion  in  two  years. The  number  of
observed  CVSO  stars (including  non-resolutions)  is  118. During  a
period of  poor seeing  we also observed  brighter stars in  Orion and
resolved J05271+0351 (HIP~25493).  The number of observations for this
program (counting repeated measurements) is 155.


\subsection{New multiple systems}
\label{sec:new}

\begin{deluxetable}{ l l r  l r   }
\tabletypesize{\scriptsize}
\tablewidth{0pt}
\tablecaption{New visual multiple systems 
\label{tab:triples}}
\tablehead{
\colhead{WDS} &
\colhead{Outer} &
\colhead{$\rho_{\rm out}$} &
\colhead{Inner} &
\colhead{$\rho_{\rm in}$} \\
& & 
\colhead{(arcsec)} & &
\colhead{(arcsec)} 
}
\startdata
03056$-$2328 &  RST 2294 AC* & 1.31  & AB & 0.60 \\
03338$-$1508 & TOK 239 AB   & 625 &  Ba,Bb*       & 0.93 \\
05441$-$1934 & HDS 766 AC*  & 1.23 & AB           & 0.07 \\
07435$+$0329 & STF 1134 AB  & 9.6  & Ba,Bb*       & 0.05 \\
08143$-$5444 & RST 3579 AB  & 0.38 & Aa,Ab*       & 0.04  \\
08159$-$3056 & BU 454  AB   & 1.86 & Ba,Bb*       & 0.35 \\
08198$-$7131 & BSO 17 AB    & 63.8 & Aa,Ab*       & 1.34  \\
08297$-$6708 & HDS 1215 AC* & 0.97 & Aa,Ab        & 0.13 \\
08429$-$7707 &  HDS1253 AB  & 0.21 & Aa,Ab*       & 0.09 \\
08515$-$8018 & LDS 244  AB  & 37.0 & Aa,Ab*       & 0.87 \\
08540$+$0825 & STT 195 AB   & 13.7 & Aa,Ab*       &  0.08  \\ 
09033$-$7036 & HEI 223  AC* & 0.46 & AB           &  0.07 \\
09173$-$6841 & I 358 AB,C   & 18.8 & Ca,Cb*       & 0.56 \\
09180$-$5453 & JNN 69 AB    & 0.52 & Aa,Ab*       & 0.05 \\
10268$-$6254 & HDS 1501 AB  & 4.0  & Ba,Bb*       & 0.09 \\
11000$-$3507 & HIP 53776AB* & 0.65 & BC*          & 0.11 \\
11155$-$6725 & HDS 1605 AC  & 2.65 & Aa,Ab*       & 0.51 \\ 
11470$-$6545 & LDS 365 AB   & 16.4 & Aa,Ab*       & 0.23 \\
12197$+$0533 & A 1597 AC*   & 1.42 & AB           & 0.67 \\
13044$-$1316 & HU 642 AC*   & 1.57 & AB           & 0.50 \\
13114$+$0938 & LDS 5771 AB  & 81.8 & Aa,Ab*       & 0.52 \\
13343$-$1132 & HDS 470 AB   & 3.76 & BC*          & 0.79 \\
13343$-$1132 & HDS 470 BC*  & 0.79 & Ca,Cb*       & 0.13 \\
14139$-$3203 & SEE 201 AB   & 17.4 & Aa,Ab*       & 0.97 \\
14243$-$6223 & RST 4525 AB  & 0.49 & BC*          & 0.07 \\
15386$-$5128 & RST 2970 AC* & 0.88 & AB           & 0.39 \\
15432$+$1340 & BU 619 AB    & 0.65 & BC*          & 0.25 \\
15495$+$2528 & WSI 111 AC*  & 0.51 & Aa,Ab        & 0.20 \\
15549$-$3731 & B 852 AB     & 0.99 & BC*          & 0.18 \\
16087$-$2523 & JNN 221 AB   & 0.82 & Aa,Ab*       & 0.05 \\
16439$-$3234 & JSP 696 AC*  & 0.93 & AB           & 0.27 \\
16509$-$1950 & B 1830 AB    & 0.39 & BC*          & 0.05 \\
16545$-$2734 & B 322 AC*    & 1.27 & AB           & 0.21 \\
18321$-$4046 & RST 4014 AB  & 0.27 & Aa,Ab*       & 0.06 \\
19243$+$2032 & HDS 2752 AC* & 0.98 & AB           & 0.27 \\ 
19251$-$2303 & RST 3225 AB  & 1.24 & Aa,Ab*       & 0.17 
\enddata 
\end{deluxetable}

As in the previous papers of this series, we report discoveries of new
visual   multiple   systems   containing   three  or   more   resolved
components. This  information is ingested into the  current version of
the  multiple-star  catalog, MSC \citep{MSC}.   Although  the high  angular
resolution  of HRCam  helps to  discover  inner close  pairs in  known
binaries,  its high  dynamic range  has also  enabled detection  of 11
faint outer  companions to known  binaries.  In HIP~53776,  both inner
and  outer pairs  are  new discoveries.   In  13343$-$1132, the  newly
discovered component  C is itself  a close pair Ca,Cb.   Resolution of
the   secondary   component    in   06401$-$3033   was   reported   by
\citet{Elliott2015},  but not  reflected in  the WDS,  so  this triple
system is {\bf re-observed} here.

Table~\ref{tab:triples} presents 35 new  multiple systems in compact
form.   Its  first column  gives  the WDS  code.   In  column (2),  the
discoverer code and the components'  designation of the outer pair are
given, followed by the separation in arcseconds in column (3). Then in
columns  (4) and  (5) the  same data are  given for  the inner
subsystem.  New  subsystems (either outer or  inner) are distinguished
by an asterisk.  Many close  inner pairs have short estimated periods,
favoring determination  of their orbits  within a few years,  like the
nearby low-mass hierarchies 09180$-$5453 and 10268$-$6254.

\begin{figure}
\epsscale{1.1}
\plotone{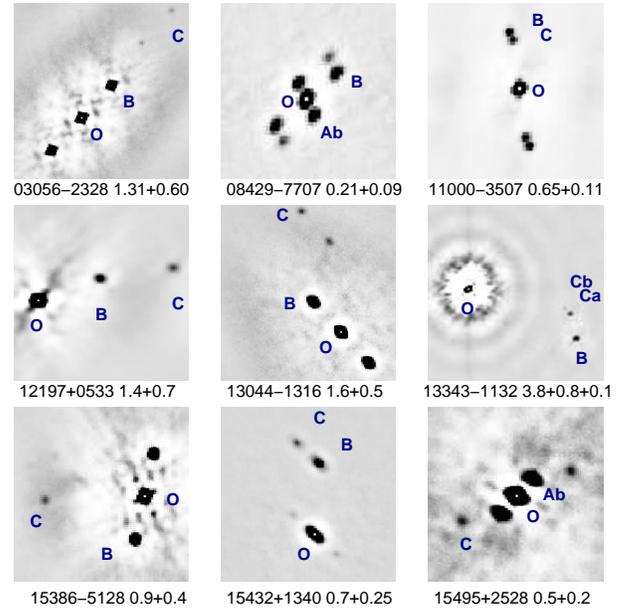}
\caption{Fragments  of  ACFs of  new  triple  systems with  comparable
  separations  between components  (trapezia).  North  up,  East left,
  scale  arbitrary. The  WDS code  and separations  in  arcseconds are
  given below each fragment;  the peaks corresponding to the locations
  of the  components are indicated  by letters; other peaks  are their
  symmetric  counterparts  and   cross-correlations  between  the  two
  secondaries (an ACF of a triple star contains 6 peaks).  The central
  peak of the ACF is marked by the white dot and the letter O.
\label{fig:triples} }
\end{figure}

In several triple systems presented here the projected outer and inner
separations are comparable  (Figure~\ref{fig:triples}). If these pairs
are  physical   and  the  true  3-dimensional   separations  are  also
comparable,  the inner and  outer orbits  strongly interact  with each
other.  Further  monitoring will help  to investigate the  dynamics of
these systems and, of course, to confirm or refute the physical nature
of the  companions.  From this  perspective, nearby systems  with fast
relative motion  will be  most interesting. On  the contrary,  a faint
tertiary  companion to  a distant  star with  a slow  PM in  a crowded
region of the sky is likely optical. Such is the case of 08297$-$6708,
15386$-$5128, 16439$-$3234, and 16545$-$2734.   The new component C in
05441$-$1934 with $\Delta I \sim 5$ mag is found in the {\it Gaia} DR2
at a slightly different position, so it is likely optical, despite the
low crowding.  The tertiary in 13044$-$1316 is likely physical because
it keeps  the same position  in DR2 while  the PM is  fast; therefore,
this triple  could be  a genuine trapezium.   The status of  other new
tertiaries remains unknown.

\subsection{New binaries}

\begin{deluxetable}{ l l c  c l  }
\tabletypesize{\scriptsize}
\tablewidth{0pt}
\tablecaption{New double stars
\label{tab:binaries}}
\tablehead{
\colhead{WDS} &
\colhead{Name} &
\colhead{$\rho$} &
\colhead{$\Delta m$} &
\colhead{Program\tablenotemark{a}} \\
 &     &    
 \colhead{(arcsec)} &
 \colhead{(mag)} & 
}
\startdata
01244$-$2803 &HIP 6566   & 1.01 & 3.7 & EH \\
01384$-$1552 &HIP 7639   & 0.51 & 5.0 & EH \\
03030$-$0205 &HIP 14194B & 0.04 & 0.0 & SB2 \\
05066$-$7734 &HIP 23776  & 0.06 & 0.5 & HIP \\ 
05271$+$0351 &HIP 25493  & 0.88 & 5.1 & Ori \\
07292$+$1246 &HIP 36371  & 0.36 & 3.2 & Ref \\
11367$-$0919 &HIP 56631  & 0.14 & 2.6 & EH \\
12120$+$0520 &HIP 59479  & 0.08 & 0.0 & EH \\
12213$-$3033 &HIP 60252  & 0.47 & 4.4 & HIP \\ 
12215$+$0749 &HIP 60272  & 0.53 & 3.2 & EH \\
12532$+$2859 &HIP 62881  & 0.10 & 0.2 & EH \\
12578$+$2252 &HIP 63262  & 1.32 & 3.2 & EH \\ 
13012$-$4109 &TDS8647CD? & 0.28 & 3.7 & MSC \\
15555$-$2616 &HIP 77984  & 0.83 & 3.8 & Ref \\
16368$+$0422 &HIP 81351  & 1.25 & 3.8 & EH \\
17111$-$2039 &HIP 84056  & 1.08 & 4.9 & Ref \\
17293$-$3839 &HIP 85583  & 1.63 & 4.2 & HIP \\
17340$-$1750 &HIP 85952  & 0.81 & 5.6 &  HIP \\
17463$-$4044 &HIP 86965  & 0.21 & 2.9 & HIP \\
17511$+$2704 &HIP 87375  & 0.42 & 3.2 & EH \\
17531$-$7501 &HIP 87539  & 0.25 & 1.9 & HIP \\
18203$-$3526 &HIP 89589  & 0.12 & 3.0 & Ref \\
18255$-$1439 &HIP 90299  & 0.20 & 0.9 &  HIP,EH \\
18393$-$3742 &HIP 91471  & 0.22 & 2.4 & HIP \\
18432$-$5730 &HIP 91808  & 0.06 & 0.0 & HIP \\
19242$-$6260 &HIP 95385  & 0.08 & 1.2 & HIP\\
19345$+$1759 &HIP 96268  & 0.07 & 0.9 & EH \\
19476$+$0105 & ENG 67Aa,Ab  & 0.62 & 5.0 & HIP \\
20446$+$1333 &HIP 102380 & 1.54 & 5.1 & EH \\
21017$-$4431 &HIP 103776 & 0.45 & 0.8 & HIP \\
22261$-$1248 &HIP 110741 & 0.73 & 3.1 & EH
\enddata 
\tablenotetext{a}{
EH -- pair resolved previously by E.H. at the WIYN telescope; 
HIP -- {\it Hipparcos} binary;
SB2 -- double-lined spectroscopic binary; 
Ref -- reference star;
Ori -- member of Orion OB1 association; 
MSC -- multiple system. }
\end{deluxetable}

In  Table~\ref{tab:binaries}  we  list  31 first-time  resolutions  of
binaries. Some of them could  have been spotted by other observers but
are not  yet published and  listed in the  WDS, being ``new''  in this
sense. Its  columns give the  WDS code, alternative  name, separation,
magnitude difference, and the  observing program code, where EH refers
to  the  list of  objects  provided by  E.H.   (all  these stars  were
resolved at  the WIYN telescope in  2012--2013), HIP is  the survey of
{\it Hiparcos}  stars, SB2 marks  double-lined spectroscopic binaries.
Four  pairs are  serendipitous  resolutions of  reference stars  (code
Ref).   The 0\farcs6  pair TDS8647CD  (13012$-$4109) belonging  to the
visual multiple system (code MSC)  is proven here to be spurious, like
many other similar {\it Tycho} binaries \citep{SAM17}, but we resolved
instead a different pair. HIP~25493  was observed as part of the Orion
OB1 survey.

\subsection{New and updated orbits}
\label{sec:orbits}.

Long  periods of classical  visual binaries  and slow  accumulation of
measures have established the  tradition of computing tentative orbits
as soon as feasible. As a result, the VB6 catalog \citep{VB6} contains
a  large number  of  provisional, low-quality  orbits. Naturally,  the
orbits  are  improved (or  drastically  revised)  in  response to  new
measures, so  that the orbit calculation becomes  an almost continuous
process.  In theory,  it could  be automated.   In  practice, however,
critical evaluation  and proper weighting of the  data (especially the
historic  visual  measures)  is  essential.   Different  authors  have
different  schemes and approaches  in this  matter.  We  adopt weights
proportional to $\sigma^{-2}$, where  the errors $\sigma$ are assigned
according  to the measurement  technique (e.g.  from 2  to 5  mas for
speckle  interferometry  at 4-m  class  telescopes,  10  mas for  {\it
  Hipparcos},  50 mas or  larger for  visual micrometer  measures) and
corrected   iteratively  to   reduce  the   impact  of   outliers,  if
necessary. The IDL program {\tt ORBIT} is used \citep{ORBIT}.

Our  speckle program  at SOAR  has contributed  to the  improvement of
existing orbits  and determination of  new orbits, especially  for the
close {\it Hipparcos} and interferometric pairs. During 2017 and 2018,
more than  200 orbits  based on the  SOAR data  were added to  the VB6
catalog.     Here    we    provide    additional   111    orbits    in
Table~\ref{tab:vborb}.  Provisional grades  and references to previous
orbits  are given  in the  last columns;  askerisks mark  orbits where
radial velocities  from the literature are used  jointly with position
measures.   For provisional  orbits of  grade 5,  we do  not  list the
formal  errors   which  are  large   and  misleading  {\bf   when  the
  observations do not  constrain all orbital elements and  we fix some
  of  them. Although  provisional  orbits are  poorly constrained,
  their  publication  helps  to  plan  further  observations  of  fast
  binaries like 04400$-$3105 (period 14 yr) and to model the motion of
  long-period  pairs,  where   no  substantial  orbit  improvement  is
  expected  in the  coming decades.   }  For  circular  and/or face-on
orbits, some  Campbell elements become  degenerate and they  are fixed
accordingly.

\LongTables
\begin{deluxetable*}{l l cccc ccc cc}    
\tabletypesize{\scriptsize}     
\tablecaption{Visual orbits
\label{tab:vborb}          }
\tablewidth{0pt}                                   
\tablehead{                                                                     
\colhead{WDS} & 
\colhead{Disc.} & 
\colhead{$P$} & 
\colhead{$T$} & 
\colhead{$e$} & 
\colhead{$a$} & 
\colhead{$\Omega$ } & 
\colhead{$\omega$ } & 
\colhead{$\i$ } & 
\colhead{Grade }  &
\colhead{Ref.\tablenotemark{a}} \\
 \colhead{{\it HIP}} &
& 
\colhead{(yr)} &
\colhead{(yr)} & &
\colhead{(arcsec)} & 
\colhead{(deg)} & 
\colhead{(deg)} & 
\colhead{(deg)} &  & 
}
\startdata
00008$+$1659 & BAG 18 & 66.62 & 1990.25 & 0.372 & 0.531 & 142.2 & -1.1 & 192.3 & 5 & new \\
00569$-$5153 & B 1418 & 19.82 & 2015.44 & 0.404 & 0.227 & 279.8 & 323.3 & 85.8 & 3 & new \\
{\it 4448} &    & $\pm$0.54 & $\pm$0.20 & $\pm$0.054 & $\pm$0.012 & $\pm$0.5 & $\pm$7.3 & $\pm$0.6&     &  \\
01205$-$1957 & TOK 203 & 8.53 & 2013.99 & 0.627 & 0.284 & 108.8 & 106.3 & 104.0 & 3 & Gln2006 \\
{\it 6273} &    & $\pm$0.37 & $\pm$0.13 & $\pm$0.044 & $\pm$0.016 & $\pm$2.0 & $\pm$2.5 & $\pm$1.9&     &  \\
01250$-$3251 & HDS 184 & 6.313 & 2018.549 & 0.515 & 0.1445 & 107.4 & 165.8 & 73.9 & 3 & new \\
{\it 6626} &    & $\pm$0.065 & $\pm$0.021 & $\pm$0.007 & $\pm$0.0010 & $\pm$0.5 & $\pm$2.1 & $\pm$0.8&     &  \\
02166$-$5026 & TOK 185 & 11.29 & 2014.17 & 0.066 & 0.090 & 270.4 & 31.8 & 45.3 & 2 & Tok2017b \\
{\it 10611} &    & $\pm$0.46 & $\pm$0.36 & $\pm$0.031 & $\pm$0.002 & $\pm$4.8 & $\pm$10.7 & $\pm$1.9&     &  \\
02254$+$0135 & HDS 315 & 600 & 2002.5 & 0.64 & 0.674 & 63.4 & 237.4 & 50.0 & 5 & new \\
02336$-$3910 & B 674 & 327.8 & 2027.35 & 0.68 & 0.238 & 36.8 & 150.1 & 172.0 & 5 & new \\
03014$+$0615 & HDS 385 & 14.893 & 2013.111 & 0.419 & 0.1169 & 161.4 & 195.5 & 54.3 & 1 & Tok2015c \\
{\it 14075} &    & $\pm$0.042 & $\pm$0.055 & $\pm$0.005 & $\pm$0.0011 & $\pm$0.9 & $\pm$1.8 & $\pm$0.7&     &  \\
03193$-$5053 & RST 70 & 51.1 & 2017.33 & 0.74 & 0.176 & 44.7 & 67.8 & 52.8 & 4 & new \\
{\it  15451} &    & $\pm$2.7 & $\pm$0.29 & $\pm$0.08 & $\pm$0.025 & $\pm$19.1 & $\pm$12.9 & $\pm$7.8&     &  \\
03305$+$2006 & RAO 11 Ba,Bb & 31.46 & 2015.18 & 0.395 & 0.294 & 63.0 & 18.8 & 103.1 & 3 & new* \\
{\it  16329} &    & $\pm$0.25 & $\pm$0.37 & $\pm$0.019 & $\pm$0.024 & $\pm$4.4 & $\pm$5.4 & $\pm$1.6&     &  \\
03363$-$1728 & HDS 456 & 21.00 & 2018.05 & 0.841 & 0.107 & 92.8 & 197.9 & 134.3 & 2 & new \\
{\it  16803} &    & $\pm$0.27 & $\pm$0.08 & $\pm$0.008 & $\pm$0.003 & $\pm$5.9 & $\pm$8.2 & $\pm$2.4&     &  \\
04028$-$3115 & HDS 511 & 84.50 & 2020.83 & 0.293 & 0.207 & 174.9 & 258.6 & 117.4 & 5 & new \\
04302$-$1747 & B 1937 & 111.55 & 1908.67 & 0.160 & 0.213 & 123.4 & 0.0 & 180.0 & 5 & Zir2008 \\
04312$+$0157 & HDS 585 & 65.1 & 2014.30 & 0.422 & 0.370 & 259.9 & 170.2 & 76.1 & 3 & new \\
{\it  21092} &    & $\pm$4.5 & $\pm$0.28 & $\pm$0.025 & $\pm$0.013 & $\pm$0.7 & $\pm$2.5 & $\pm$0.7&     &  \\
04375$+$1509 & CHR 153 & 127.7 & 2047.35 & 0.251 & 0.712 & 144.6 & 75.1 & 74.9 & 5 & new \\
04400$-$3105 & HDS 602 & 14.0 & 1995.58 & 0.700 & 0.176 & 108.2 & 75.3 & 139.0 & 5 & new \\
04518$+$1339 & BU 552  AB & 97.7 & 1982.193 & 0.592 & 0.7432 & 142.6 & 312.3 & 50.3 & 2 & Sod1999 \\
{\it  22607} &    & $\pm$1.4 & $\pm$0.054 & $\pm$0.003 & $\pm$0.0066 & $\pm$0.6 & $\pm$0.4 & $\pm$0.3&     &  \\
05048$+$1319 & HEI 104 & 140 & 2034.124 & 0.420 & 0.160 & 182.4 & 139.9 & 84.6 & 5 & new \\
05103$-$0736 & A 484 & 18.936 & 2000.147 & 0.787 & 0.1555 & 111.5 & 309.9 & 106.4 & 2 & Tok2017b* \\
{\it 24076}  &    & $\pm$0.064 & $\pm$0.065 & $\pm$0.005 & $\pm$0.0019 & $\pm$0.5 & $\pm$0.9 & $\pm$0.5&     &  \\
05267$-$6436 & I 1150 & 907 & 2020.71 & 0.822 & 0.722 & 171.5 & 212.0 & 55.5 & 5 & new \\
05334$-$4923 & HDS 732  Aa,Ab & 21.9 & 2020.609 & 0.850 & 0.162 & 313.1 & 245.2 & 76.0 & 4 & new \\
{\it  26050} &    & $\pm$1.2 & $\pm$1.445 & fixed & $\pm$0.037 & $\pm$15.0 & $\pm$4.9 & $\pm$6.9&     &  \\
05427$-$6708 & I 745 & 208 & 2017.19 & 0.775 & 0.568 & 238.3 & 205.8 & 72.1 & 4 & new \\
{\it  26904} &    & $\pm$33 & $\pm$0.29 & $\pm$0.021 & $\pm$0.051 & $\pm$2.2 & $\pm$2.4 & $\pm$1.4&     &  \\
05505$-$0310 & HDS 785 & 105.6 & 2021.9 & 0.65 & 0.208  & 170.2 & 147.9 & 118.0 & 5 & new \\
05590$-$0740 & HDS 809 & 100 & 216.13 & 0.314 & 0.457 & 29.9 & 129.9 & 31.5 & 5 & new \\
06023$+$0142 & CHR 162 & 200 & 2033.556 & 0.88 & 0.3242 & 219.4 & 238.2 & 102.8 & 5 & new \\
06143$-$1729 & A 3025 & 226 & 2016.09 & 0.850 & 0.588 & 94.9 & 128.7 & 154.7 & 4 & new \\
{\it  29601} &    & $\pm$17 & $\pm$0.04 & $\pm$0.008 & $\pm$0.027 & $\pm$11.6 & $\pm$13.0 & $\pm$4.8&     &  \\
06201$-$0752 & HDS 866 & 58.23 & 2023.03 & 0.66 & 0.186 & 82.6 & 243.7 & 81.4 & 5 & new \\
06314$+$0749 & A 2817 & 31.71 & 2015.333 & 0.290 & 0.1943 & 55.2 & 142.0 & 38.4 & 1 & Tok2015c \\
{\it  31089} &    & $\pm$0.19 & $\pm$0.075 & $\pm$0.004 & $\pm$0.0012 & $\pm$1.4 & $\pm$2.3 & $\pm$0.9&     &  \\
06510$+$0551 & HDS 950 & 30.5 & 2016.59 & 0.718 & 0.106 & 165.9 & 242.4 & 134.5 & 5 & new \\
06533$-$1902 & CHR 169  & 37.37 & 2017.80 & 0.487 & 0.186 & 174.5 & 75.7 & 107.2 & 3 & Tok2017b \\
{\it  33077} &    & $\pm$0.60 & $\pm$0.08 & $\pm$0.020 & $\pm$0.003 & $\pm$0.9 & $\pm$1.2 & $\pm$1.0&     &  \\
06584$-$1300 & HDS 969 AB & 68.4 & 1979.1 & 0.164 & 0.795 & 38.3 & 217.6 & 93.7 & 5 & new \\
07003$-$2207 & FIN 334 Aa,Ab & 217.5 & 2031.7 & 0.072 & 0.142 & 140.5 & 180.0 & 111.1 & 5 & Doc2018e \\
07040$-$4337 & TOK 390 Ca,Cb & 4.623 & 2011.808 & 0.462 & 0.159 & 58.7 & 232.0 & 161.7 & 3 & SaJ2011 \\
{\it  34069} &    & $\pm$0.040 & $\pm$0.060 & $\pm$0.014 & $\pm$0.004 & $\pm$21.0 & $\pm$19.5 & $\pm$8.1&     &  \\
07116$-$7959 & HDS 998 & 61.6 & 2021.5 & 0.095 & 0.117 & 34.7 & 53.6 & 40.5 & 5 & new \\
07167$+$1609 & HDS 1007 & 28.22 & 2014.51 & 0.319 & 0.228 & 345.4 & 126.5 & 81.9 & 3 & new \\
{\it 35219}  &    & $\pm$0.29 & $\pm$0.13 & $\pm$0.011 & $\pm$0.002 & $\pm$0.4 & $\pm$2.3 & $\pm$0.6&     &  \\
07336$+$1550 & MCA 32 & 167 & 1993.71 & 0.879 & 0.428 & 97.1 & 269.1 & 83.6 & 4 & Zir2008 \\
{\it  36760} &    & $\pm$27 & $\pm$0.17 & $\pm$0.033 & $\pm$0.058 & $\pm$0.4 & $\pm$0.8 & $\pm$0.7&     &  \\
07427$-$3510 & HDS 1091 & 120.6 & 2019.9 & 0.50 & 0.190 & 89.6 & 216.2 & 118.2 & 5 & new \\ 
08342$-$0957 & HDS 1226 & 55 & 2003.06 & 0.398 & 0.203 & 201.5 & 112.0 & 118.6 & 5 & new \\
08444$-$4428 & HDS 1256 & 12.695 & 2011.75 & 0.536 & 0.2902 & 140.5 & 60.3 & 146.9 & 4 & new \\
{\it  42881} &    & $\pm$0.068 & $\pm$0.60 & $\pm$0.091 & $\pm$0.0359 & $\pm$21.6 & $\pm$29.6 & $\pm$13.5&     &  \\
08454$-$0013 & A 2548 & 1100 & 1982.7 & 0.575 & 0.4974 & 179.2 & 230.1 & 114.9 & 5 & new \\
08476$-$3124 & HDS 1273 & 84.3 & 2020.38 & 0.572 & 0.272 & 236.5 & 90.0 & 99.4 & 5 & new \\
08514$-$5047 & HDS1281 & 25.8 & 2022.2 & 0.50 & 0.133 & 40.8 & 118.3 & 140.9 & 5 & new \\
08571$+$1139 & HDS1296 & 34.22 & 2003.33 & 0.836 & 0.580 & 219.2 & 131.2 & 105.6 & 3 & new \\
{\it  43948} &    & $\pm$0.59 & $\pm$0.07 & $\pm$0.006 & $\pm$0.010 & $\pm$0.4 & $\pm$1.2 & $\pm$0.4&     &  \\
09024$-$6624 & TOK 197 & 0.652 & 2015.593 & 0.041 & 0.0321 & 105.3 & 248.5 & 101.9 & 3 & Tok2018i \\
{\it   44382} &    & $\pm$0.001 & $\pm$0.063 & $\pm$0.038 & $\pm$0.0013 & $\pm$1.4 & $\pm$34.2 & $\pm$2.3&     &  \\
09086$-$2550 & TOK 357 BC & 60 & 2070.0 & 0.20 & 0.299 & 177.0 & 39.6 & 66.8 & 5 & new \\
09100$-$2845 & B 179 & 80.35 & 2026.88 & 0.562 & 0.372 & 169.4 & 158.5 & 116.3 & 4 & Doc2013c \\
{\it 45003}  &    & $\pm$0.84 & $\pm$0.62 & $\pm$0.025 & $\pm$0.006 & $\pm$1.1 & $\pm$3.0 & $\pm$1.3&     &  \\
09293$-$4432 & HDS 1360 Aa,Ab & 79.5  & 2031.8  & 0.82  & 0.661  & 56.1 & 66.7 & 115.3 & 5 & new \\
09442$-$2746 & FIN 326 & 18.394 & 2020.96 & 0.504 & 0.107 & 175.3 & 138.9 & 127.0 & 2 & Doc2013d \\
{\it  47758} &    & $\pm$0.088 & $\pm$0.17 & $\pm$0.019 & $\pm$0.002 & $\pm$2.3 & $\pm$4.1 & $\pm$1.1&     &  \\
09522$+$0807 & A 2762 & 640 & 2092.6 & 0.60 & 1.603 & 131.1 & 17.3 & 105.8 & 5 & new \\
09535$+$1657 & CHR 219 & 54.0 & 2023.1 & 0.274 & 0.308 & 243.0 & 245.7 & 105.2 & 3 & Hrt2012a \\
{\it  48504} &    & $\pm$9.4 & $\pm$1.7 & $\pm$0.056 & $\pm$0.020 & $\pm$1.8 & $\pm$23.8 & $\pm$0.9&     &  \\
10067$+$1754 & HDS 1457 & 203 & 1989.2 & 0.649 & 0.684 & 109.1 & 298.9 & 126.2 & 5 & MaB2016 \\
10174$-$5354 & CVN 16 Aa,Ab & 5.327 & 2005.936 & 0.139 & 0.0952 & 129.0 & 96.5 & 15.3 & 2 & Cvn2009 \\
 \ldots &    & $\pm$0.021 & $\pm$0.067 & $\pm$0.013 & $\pm$0.0023 & $\pm$25.9 & $\pm$25.7 & $\pm$7.4&     &  \\
10214$-$2616 & HDS 1491 & 22.1 & 2022.75 & 0.287 & 0.113 & 262.8 & 0.0 & 180.0 & 5 & new \\
10260$+$0256 & A 2570 & 174 & 2023.5 & 0.83 & 0.245 & 122.9 & 329.4 & 114.3 & 5 & Zir2014a \\
10264$+$2545 & HDS 1500 & 85 & 1978.67 & 0.136 & 0.2130 & 150.8 & 317.6 & 65.2 & 5 & new \\
10388$-$4245 & FIN 338 & 80     & 2021.7  & 0.454 & 0.156 & 37.1 & 187.6 & 97.0 & 4 & new \\
{\it  52112} &    & fixed & $\pm$2.1  & $\pm$0.041 & $\pm$0.008 & $\pm$0.8 & $\pm$8.4 & $\pm$2.0&     &  \\
10419$-$7811 & HDS 1530 & 39.07 & 2007.88 & 0.547 & 0.267 & 109.7 & 132.6 & 50.2 & 3 & Tok2015c \\
{\it  52351} &    & $\pm$0.67 & $\pm$0.10 & $\pm$0.008 & $\pm$0.004 & $\pm$1.5 & $\pm$1.1 & $\pm$0.7&     &  \\
10455$-$2502 & I 502 AB & 256 & 2016.58 & 0.728 & 0.307 & 225.2 & 0.0 & 180.0 & 4 & new \\
{\it 52615}  &    & $\pm$44 & $\pm$0.22 & $\pm$0.028 & $\pm$0.028 & $\pm$2.4 & fixed & fixed&     &  \\
10479$-$6416 & HDS 1544 & 78 & 2003.8 & 0.319 & 0.224 & 96.3 & 175.0 & 127.5 & 5 & new \\
11014$-$1204 & HDS 1572 & 18.45 & 2013.69 & 0.682 & 0.172 & 142.4 & 133.6 & 97.4 & 3 & Tok2015c \\
{\it  53879} &    & $\pm$0.70 & $\pm$0.03 & $\pm$0.014 & $\pm$0.003 & $\pm$0.4 & $\pm$1.4 & $\pm$0.5&     &  \\
11151$-$3929 & SEE 128 & 95.0 & 1988.4 & 0.52 & 0.1353 & 165.1 & 72.9 & 46.4 & 3 & new \\
{\it  54949} &     & $\pm$4.5 & $\pm$1.5 & $\pm$0.06 & $\pm$0.0161 & $\pm$11.5 & $\pm$13.9 & $\pm$7.9&     &  \\
11250$-$3200 & CHR 242 Aa,Ab & 13.47 & 2010.74 & 0.563 & 0.137 & 123.7 & 211.0 & 115.7 & 3 & new \\
{\it  55714} &    & $\pm$0.13 & $\pm$0.14 & $\pm$0.020 & $\pm$0.002 & $\pm$1.7 & $\pm$4.7 & $\pm$1.0&     &  \\
11272$-$1604 & HDS 1627 Aa,Ab & 46.3 & 2003.5 & 0.358 & 0.222 & 89.9 & 165.0 & 119.5 & 4 & new \\
{\it 55884}  &    & $\pm$3.2 & $\pm$3.1 & $\pm$0.031 & $\pm$0.026 & $\pm$4.7 & $\pm$22.3 & $\pm$7.0 &     &  \\
12096$-$6727 & HDS 1716 & 72  & 2018.17 & 0.525 & 0.157 & 58.8 &333.9& 40.4 & 5 & new \\
12250$-$0414 & TOK 400 & 21.1 & 2018.77 & 0.61 & 0.211 & 96.9 & 182.0 & 125.9 & 5 & new \\
12419$-$6444 & HDS 1779 & 49 & 2019.86 & 0.675      & 0.124 & 206.7 & 128.2 & 118.1 & 4 & new \\
{\it  61959} &    & $\pm$15& $\pm$0.20 & $\pm$0.10  & $\pm$0.031 & $\pm$3.1 & $\pm$10.5 & $\pm$10.5&     &  \\
13081$-$7719 & HDS 1839 & 17.01 & 2015.01  & 0.160 & 0.205 & 167.8 & 163.6 & 119.5 & 3 & Tok2016e \\
{\it  64091} &    & $\pm$0.36 & $\pm$0.36  & $\pm$0.026 & $\pm$0.005 & $\pm$1.5 & $\pm$7.6  & $\pm$1.2&     &  \\
13417$-$2915 & HDS 1922 & 88.4 & 2003.08 & 0.56 & 0.215 & 257.2 & 72.7 & 110.4 &  5 & new \\
14094$+$1015 & RAO 16 & 8.36 & 2018.88 & 0.98 & 0.104 & 114.0 & 66.0 & 114.4 & 5 & new \\
14261$-$6536 & HDS 2031 & 31.34 & 2009.26 & 0.50 & 0.183 & 245.1 & 133.3 & 104.2 & 5 & new \\
14383$-$4954 & FIN 371 & 51.6 & 2015.5 & 0.201 & 0.099 & 233.8 & 38.0 & 108.0 & 3 & Tok2016e \\
{\it  71577} &    & $\pm$3.3 & $\pm$1.3 & $\pm$0.027 & $\pm$0.004 & $\pm$1.3 & $\pm$10.7 & $\pm$1.1&     &  \\
15481$-$2513 & HDS 2226 & 31.1 & 2010.38 & 0.499 & 0.106 & 57.7 & 204.7  & 140.8 & 4 & new \\
{\it 77399}  &    & $\pm$1.0 & $\pm$0.60 & $\pm$0.159 & $\pm$0.008 & $\pm$11.7& $\pm$20.1 & $\pm$19.4&     &  \\
15544$-$6131 & HDS 2240 & 80 & 2020.57 & 0.83 & 0.188 &183.4 & 0.0 6 & 0.0  & 5 & new \\
16115$+$0943 & FIN 354 & 61.1 & 1999.62 & 0.066 & 0.1281 & 263.8 & 91.6 & 90.2 & 3 & Doc2013d \\
{\it  79337} &    & $\pm$1.7 & $\pm$0.71 & $\pm$0.047 & $\pm$0.0009 & $\pm$0.3 & $\pm$4.3 & $\pm$0.6&     &  \\
16115$+$0943 & FIN 354 & 29.68 & 2000.91 & 0.742 & 0.0751 & 84.2 & 196.9 & 90.5 & 3 & Doc2013d \\
{\it  79337} &    & $\pm$0.22 & $\pm$2.34 & $\pm$0.124 & $\pm$0.0039 & $\pm$0.4 & $\pm$27.0 & $\pm$1.3&     &  \\
16143$-$1025 & RST 3936 AB & 35 & 1999.92 & 0.818 & 0.173 & 263.6 & 160.6 & 109.7 & 5 & new \\
16161$-$3037 & I 1586 & 160 & 2050.85 & 0.200 & 0.361 & 0.1 & 261.2 & 137.9 & 5 & new \\
16385$-$5728 & TOK 51 Aa,Ab & 25 & 2027.73 & 0.328 & 0.262 & 59.9 & 206.9 & 95.8 & 5 & new \\
16514$-$2450 & B 2397 & 69.4 & 2021.8 & 0.043 & 0.1534 & 22.4 & 76.6 & 120.5 & 3 & new \\
{\it  82474} &    & $\pm$3.4 & $\pm$9.3 & $\pm$0.034 & $\pm$0.0024 & $\pm$2.9 & $\pm$52.0 & $\pm$1.6&     &  \\
17309$-$5621 & FIN 257 & 700 & 2017.95 & 0.80 & 0.790 & 55.7 & 49.2 & 49.2 & 5 & new \\
17430$+$0547 & HDS 2506 & 21.07 & 2007.40 & 0.553 & 0.374 & 129.5 & 237.0 & 104.3 & 3 & new \\
{\it  86707} &    & $\pm$0.26 & $\pm$0.03 & $\pm$0.008 & $\pm$0.004 & $\pm$0.5 & $\pm$1.2 & $\pm$0.4&     &  \\
17541$-$4821 & B 1870 & 111.3 & 1999.76 & 0.487 & 0.153 & 110.7 & 151.6 & 126.9 & 3 & new \\ 
{\it  87635} &    & $\pm$6.7  & $\pm$0.75 & $\pm$0.024 & $\pm$0.008 & $\pm$3.5 & $\pm$2.4 & $\pm$2.3&     &  \\
17577$-$2143 & HDS 2530 & 59.9 & 1998.25 & 0.608 & 0.514 & 143.7 & 250.9 & 58.9 & 4 & new \\
{\it  87925} &    & $\pm$2.1 & $\pm$0.56 & $\pm$0.019 & $\pm$0.013 & $\pm$2.6 & $\pm$2.3 & $\pm$1.6&     &  \\
18078$+$2606 & CHR 67 Aa,Ab & 35.53 & 2002.19 & 0.100 & 0.2954 & 144.8 & 82.7 & 77.1 & 2 & Msn2001a \\
{\it  88818} &    & $\pm$0.16 & $\pm$0.26 & fixed & $\pm$0.0026 & $\pm$0.5 & $\pm$2.4 & $\pm$0.5&     &  \\
18166$-$2033 & MCA 51 & 119 & 2031.15 & 0.50 & 0.179 & 133.7 & 164.3 & 91.0 & 4 & new \\
{\it  89567} &    & $\pm$56 & $\pm$8.12 & fixed & $\pm$0.028 & $\pm$0.6 & $\pm$67.8 & $\pm$0.6&     &  \\
18448$-$3323 & OL 20 & 490 & 2030.6 & 0.46 & 0.607 & 162.4 & 304.0 & 109.0 & 5 & new \\
18520$-$5418 & TOK 325 Aa,Ab & 13.2 & 2017.12 & 0.314 & 0.106 & 110.7 & 308.3 & 47.3 & 3 & Tok2017b \\
{\it  92592} &    & $\pm$1.8 & $\pm$0.21 & $\pm$0.064 & $\pm$0.004 & $\pm$4.7 & $\pm$13.0 & $\pm$5.6&     &  \\
19029$-$5413 & I 1390 & 47.6 & 2009.434 & 0.666 & 0.185 & 73.8 & 220.7 & 58.6 & 3 & Tok2015c \\
{\it  93524} &    & $\pm$1.8 & $\pm$0.047 & $\pm$0.010 & $\pm$0.004 & $\pm$1.0 & $\pm$1.5 & $\pm$1.4&     &  \\
19117$-$2604 & RST 2094 & 245 & 2046.7  & 0.556 & 0.709 & 43.6 & 195.5 & 75.3 & 5 & new \\
19194$-$0136 & HDS 2734 Aa,Ab & 36.83 & 2020.10 & 0.625 & 0.296 & 19.85 & 0 & 0 & 3 & Tok2015c \\
{\it 94960}  &    & $\pm$0.29 & $\pm$0.03 & $\pm$0.012 & $\pm$0.001 & $\pm$0.40 & fixed & fixed &     &  \\
19240$-$5320 & HDS 2751 & 20.61 & 2019.52 & 0.60 & 0.119 & 157.4 & 67.0 & 33.6 & 4 & new \\
{\it  95360} &    & $\pm$0.61 & $\pm$0.13 & fixed & $\pm$0.006 & $\pm$12.8 & $\pm$9.6 & $\pm$5.2&     &  \\
19407$-$0037 & CHR 88 Aa,Ab & 10.155 & 2012.847 & 0.463 & 0.0631 & 191.2 & 0.0 & 180.0 & 2 & Tok2015c \\
{\it  96807} &    & $\pm$0.033 & $\pm$0.056 & $\pm$0.010 & $\pm$0.0006 & $\pm$1.1 & fixed & fixed&     &  \\
19453$-$6823 & TOK 425 Ba,Bb & 4.12 & 2017.09 & 0.80 & 0.0504 & 136.6 & 202.0 & 124.5 & 4 & new \\
{\it  97196} &    & $\pm$0.24 & $\pm$0.12 & fixed & $\pm$0.0053 & $\pm$12.8 & $\pm$24.6 & $\pm$8.3&     &  \\
19531$-$1436 & CHR 90 & 343.2 & 1998.85 & 0.716 & 0.678 & 3.3 & 73.5 & 126.5 & 5 & Cve2010b \\
21206$+$1310 & HDS 3038 & 78.2  & 2012.84 & 0.25 & 0.231 & 126.1 &274.3 & 91.1 & 5 & new \\
21330$+$2408 & HDS 3065 Aa,Ab & 67.3 & 2025.0 & 0.63 & 0.466 & 75.8 & 30.7 & 101.3 & 5 & new \\
21522$+$0538 & JOD 23 AB & 9.39 & 2019.48 & 0.417 & 0.142 & 161.9 & 0.0 & 0.0 & 4 & new \\
{\it 107948} &    & $\pm$0.13 & $\pm$0.04 & $\pm$0.011 & $\pm$0.002 & $\pm$1.3 & fixed & fixed&     &  \\
22003$-$2330 & I 674 & 301 & 2007.87 & 0.852 & 0.500 & 69.4 & 44.5 & 59.7 & 3 & new \\
 \ldots &    & $\pm$57 & $\pm$0.11 & $\pm$0.019 & $\pm$0.059 & $\pm$1.2 & $\pm$1.8 & $\pm$1.1&     &  \\
22061$-$0521 & TOK 373 & 13.1 & 2013.53 & 0.366 & 0.182 & 45.6 & 352.3 & 96.4 & 4 & new \\
{\it 109110} &    & fixed & $\pm$0.22 & $\pm$0.021 & $\pm$0.003 & $\pm$0.7 & $\pm$8.0 & $\pm$1.1&     &  \\
22083$+$2409 & HDS 3145 & 10.641 & 1997.760 & 0.516 & 0.0951 & 62.0 & 301.9 & 149.4 & 1 & Bag2007b \\
{\it 109281} &    & $\pm$0.044 & $\pm$0.048 & $\pm$0.013 & $\pm$0.0012 & $\pm$4.5 & $\pm$4.3 & $\pm$2.2&     &  \\
22116$-$3428 & CHR 230 Aa,Ab & 43.4 & 2010.29 & 0.831 & 0.104 & 130.7 & 134.1 & 75.1 & 3 & Tok2016e \\
{\it 109561} &    & $\pm$2.4 & $\pm$0.37 & $\pm$0.023 & $\pm$0.0010 & $\pm$2.0 & $\pm$6.4 & $\pm$2.1&     &  \\
22126$-$1802 & HDS 3153 & 35.7 & 2022.09 & 0.78 & 0.177 & 57.2 & 49.1 & 124.2 & 5 & new \\
22259$-$7501 & TOK 434 Ba,Bb & 11.0 & 2022.21 & 0.25 & 0.254 & 235.0 & 270.0 & 87.3 & 5 & new \\
22357$-$2808 & HDS 3208 Aa,Ab & 19.13 & 2021.76 & 0.51 & 0.174 & 133.6 & 209.3 & 70.9 & 4 & new \\
{\it 111520} &    & $\pm$0.95 & $\pm$1.32 & $\pm$0.11 & $\pm$0.016 & $\pm$3.3 & $\pm$12.5 & $\pm$4.8&     &  \\
22376$+$2400 & HDS 3212 & 31.22 & 2027.60 & 0.357 & 0.110 & 133.6 & 145.7 & 31.1 & 4 & new \\
{\it 111694} &    & $\pm$0.58 & $\pm$1.01 & $\pm$0.021 & $\pm$0.003 & $\pm$9.6 & $\pm$14.6 & $\pm$7.8&     &  \\
22493$+$1517 & HDS 3241 & 92.5 & 2005.6 & 0.703 & 0.235 & 144.9 & 348.7 & 52.6 & 3 & Doc2013f \\
{\it 112695} &    & $\pm$12.9 & $\pm$0.1 & $\pm$0.028 & $\pm$0.022 & $\pm$1.6 & $\pm$1.6 & $\pm$1.6&     &  \\
23270$-$1515 & HU 297 & 92.6 & 1983.5 & 0.90 & 0.374 & 142.5 & 48.6 & 111.5 & 4 & new \\
{\it 115742} &    & $\pm$2.9 & $\pm$1.8 & fixed & $\pm$0.067 & $\pm$3.2 & $\pm$11.3 & $\pm$7.8&     &  \\
23286$-$3821 & HDS 3342 & 46.6  & 2014.9  & 0.368 & 0.119  & 125.3 & 155.6 & 122.8 & 5 & new \\
23350$+$0136 & MEL 9 BC & 31.81 & 1998.66 & 0.0 & 0.437 & 10.5 & 0.0 & 84.2 & 4 & new \\
{\it 116384} &    & $\pm$0.18 & $\pm$0.06 & fixed & $\pm$0.005 & $\pm$0.6 & fixed & $\pm$0.5&     &  \\
23597$-$4405 & WSI 140 & 11.77 & 2001.98 & 0.394 & 0.217 & 93.4 & 0.0 & 0.0 & 3 & TSN2017    \\
\ldots &    & $\pm$0.39 & $\pm$0.39 & $\pm$0.007 & $\pm$0.004 & $\pm$2.6 & fixed & fixed &     &  
\enddata 
\tablenotetext{a}{References to VB6 are provided at
  \url{http://ad.usno.navy.mil/wds/orb6/wdsref.txt} }
\end{deluxetable*}

\begin{figure}
\epsscale{1.1}
\plotone{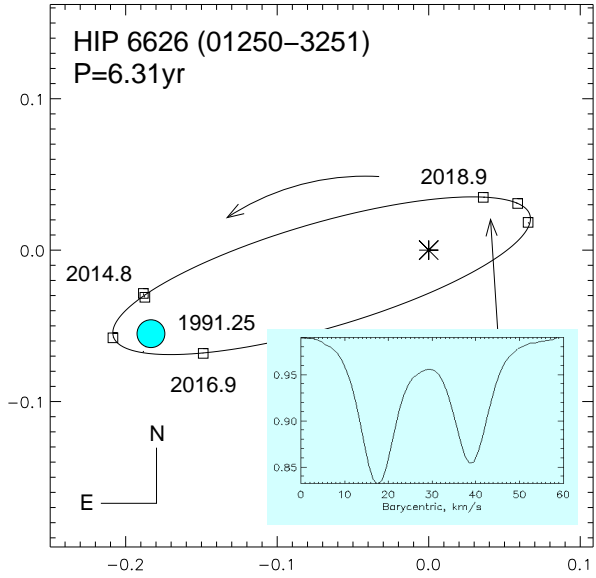}
\caption{Orbit of  HIP 6626 (HDS  184, WDS J01250-3251).   The primary
  component is located at the coordinate origin, with scale in arcseconds.
  The  ellipse marks  the  trajectory.  The  original {\it  Hipparcos}
  discovery  is marked  by the  blue  circle, the  SOAR measures  from
  2014.8 to 2018.9 by squares.  The insert shows double lines observed
  on 2018.916, shortly after the periastron passage.
\label{fig:HIP6626} }
\end{figure}

As an example  of a ``fast'' binary from  our {\it Hipparcos} program,
we  show  in  Figure~\ref{fig:HIP6626}  the first  orbit  of  HIP~6626
(HDS~184, GJ 1083), a K7V dwarf within 25\,pc from the Sun. Measurements
at  SOAR taken during  four  years,  together with  the  first {\it  Hipparcos}
resolution, define the orbit quite well. The short period of 6.3
years implies a large  radial velocity (RV) amplitude. Realizing this,
we took one spectrum with CHIRON on 2018.916 and, indeed, detected the
double  lines  with  an  RV  difference of  20  km~s$^{-1}$.   Further
monitoring and accurate parallax  from future {\it Gaia} data releases
will lead to precise mass measurement of these stars.

\begin{figure}
\epsscale{1.1}
\plotone{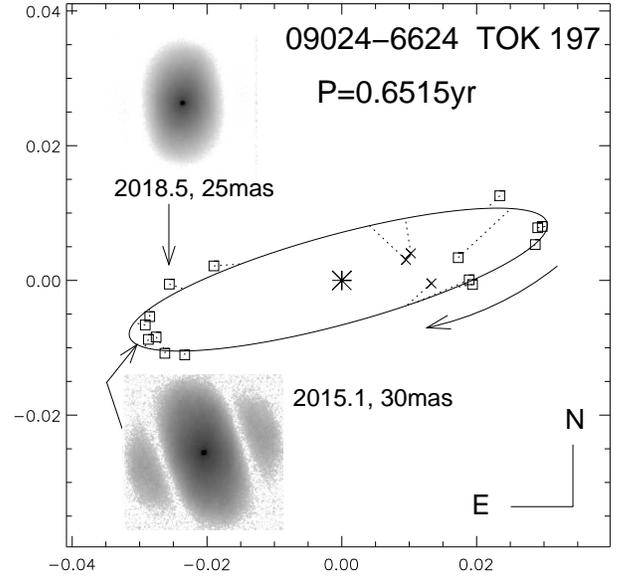}
\caption{Orbit of $\alpha$~Vol (WDS J09024-6624). The inserts show
  power spectra recorded in 2015.1 near maximum separation and in
  2018.5 at separation of 25\,mas. Crosses denote non-resolutions.
\label{fig:TOK197} }
\end{figure}

Figure~\ref{fig:TOK197} illustrates  a particularly difficult  case of
orbit calculation.  The 4th magnitude star $\alpha$ Volantis (HR~3615,
HD~78045,  HIP~4438, spectral  type kA3hA5mA5V)  has been  resolved at
SOAR  in  2010 at  29\,mas  and later  measured  17  times at  similar
separations or unresolved.  It was  placed on the observing program on
request by  J.~Patience, as part of  the survey of  Herbig AeBe stars.
With the  small separation, small $\Delta  y = 0.2$  mag, and frequent
non-resolutions,  it was  difficult  to make  sense  of the  available
measures.   In the  beginning  of  2018, a  provisional  orbit with  a
one-year period  was computed.  To test  it, the star  was observed in
2018.5,   outside  its  normal   visibility  season.    This  critical
observation invalidated  the proposed  orbit, but helped  to establish
the  true orbital period  of 0.6515$\pm$0.001  years (238  days).  All
measures  were  examined and  re-processed  where necessary,  reaching
below the nominal diffraction  limit of 30\,mas and down-weighting the
data affected  by telescope vibration.  The weighted  rms residuals to
the orbit  are 1.5 mas in  both coordinates.  With the  {\it Gaia} DR2
parallax  of 26.49  mas, the  mass  sum is  4.2 solar.   Based on  its
kinematics, the star may belong to the 300-Myr old UMa moving group.

Several binaries in  Table~\ref{tab:vborb} have sub-solar metallicity.
For example,  HIP~24076 (05103$-$0736, A~484)  with [Fe/H]=$-$0.57 dex
\citep{Holmberg2009}   goes  through  the   periastron  of  its
eccentric orbit in 2019.0 and is being followed both by speckle and by
spectroscopy. Accurate orbits and masses  will be used to test stellar
models, continuing the work of \citet{Horch2019} on metal-poor stars.

\subsection{Spurious double stars}
\label{sec:bogus}

A star is considered to be double if it was resolved at least once. If
the resolution was spurious,  as shown by subsequent observations, the
double-star  label still  persists. It  is difficult  to prove  that a
given  star is  {\it not  double} because  its non-resolutions  can be
explained by the  orbital motion that brings the  components too close
together, by a large $\Delta m$, or by poor observing conditions.  The
WDS  records  only  the  last  measure,  so, when  a  given  pair  was
repeatedly unconfirmed,  this fact is  hidden and, instead,  leaves an
impression that the object is ``neglected'' by  observers. Here we
present  a list  of likely  spurious pairs,  hoping to  clean  the WDS
catalog and to reduce the waste  of effort for their observation. In a
sense, this is a necessary complement of new discoveries presented
above.

Two enigmatic  cases of ``ghost'' visual pairs  with multiple spurious
historic  measures were  presented by  \citet{Tok2012a};  other likely
spurious visual  binaries are suggested here, namely  several pairs by
van  den  Bos  (discoverer  code   B).   An  intriguing  case  is  WDS
J03244$-$1539  (A~2909AB), for which  a grade  3 orbit  with $P=11.35$
years was computed.  This object  was visited at SOAR 13 times between
2007 and 2018  and resolved only once in  2013.74;  the non-resolutions
contradict the orbit, and we believe  that this star is single (it has
a  constant RV).   Other observing  techniques also  contributed their
share   of  spurious   pairs,   for  various   reasons.   In   speckle
interferometry,  doubling or  elongation  can be  caused by  telescope
vibration,  optical ghosts  \citep[see][]{SAM17}, or  poorly corrected
atmospheric  dispersion.  A number  of CHARA  pairs were  withdrawn as
false resolutions  by \citet{McA1993}; several more  are spotted here.
Similarly, some  resolutions at SOAR (discoverer code  TOK) are likely
spurious, as revealed  by subsequent observations.  Lunar occultations
have supplied  quite a few false  doubles stars, and  many {\it Tycho}
pairs  with  small separations  are  spurious  as well  \citep{SAM17}.
Other  reasons of  spurious discoveries  are optical  pairs  with fast
relative motion and pointing wrong stars.

It is almost  impossible to prove with certainty that  a given star is
not double; our conclusions on  the spurious nature  of some pairs
are based on the available evidence.  When a pair discovered visually is
repeatedly unresolved with a  more powerful technique such as speckle,
it is very likely spurious.  Estimation of the orbital period based on
angular  separation and  distance from the Sun helps  to reject  the pair  when its
speckle coverage is of comparable  duration or when the period is very
long, as the  usual hypothesis that the binary  became temporarily too
close can be dropped.  A number of spurious subsystems can be rejected
because the outer binaries were repeatedly measured by speckle without
resolving the subsystem, as, for example, WDS J15462$-$280, one of our
calibrators;   its    subsystem   CHR~50   is    definitely   spurious
\citep{Tok2012a}.

\begin{deluxetable}{ l l l l  } 
\tabletypesize{\scriptsize}    
\tablecaption{Spurious pairs
\label{tab:bogus} }                   
\tablewidth{0pt}     
\tablehead{ \colhead{WDS}  &
\colhead{Discoverer}  &  
\colhead{Resolved} & 
\colhead{Unresolved\tablenotemark{a}} 
}
\startdata  
00023$-$2943 & B 631 & 3\arcsec Vis 1925-27 & 2018, DR2 \\ 
00028$-$2353 & B 632  & 0\farcs2  Vis 1926-31 & 2018, L \\ 
02098$-$4052 & TOK 427 & 0\farcs4 Sp 2014 & Vib \\ 
03244$-$1539 & A 2909AB & 0\farcs1 Vis 1918-2013 & 2007--18, S \\  
03590$-$0056 & HEI 215AB & 1\farcs8 Vis 1973-97  & 2018, L \\ 
06273$+$1453 & CHR 251 & 0\farcs05 Sp 1995 & 2016--18, L \\ 
06448$-$0424 & HDS 937   &  0\farcs5 HIP & 2016--18, L \\    
06461$-$2045 & I 760 & 1\farcs2 Vis 1910 & 2018, DR2 \\ 
06523$-$0510 &  WSI 125Bab & 0\farcs1 Sp. 2010 & 2014--17, DR2 \\ 
06533$-$1528 & HDS 954 & 0\farcs6 HIP & 2018, L \\ 
06585$$-$$2406 & HDS 971  & 1\farcs0 HIP & 2015-18, L  \\ 
07185$-$5724 & RST 244Bab & 0\farcs9 Sp 2010-16   & Alias \\ 
07431$+$0011 & B 2526AB & 0\farcs1 Vis 1936-62 & 1976--2018 \\ 
07501$-$2815 & HDS1113 & 0\farcs4 HIP & 2015-18, L \\ 
08095$-$4720 & WSI 55Bab & 0\farcs1 Sp 2006-09 & 2014--18, L \\ 
08107$-$7430 & B 1981AB & 0\farcs2 Vis 1936 & 2018, DR2 \\ 
09128$-$6055 & CHR 144Aab & 0\farcs02 Sp 1989 & 1990--2018, S \\ 
10311$-$2411 & CHR 132Aab & 0\farcs1 Sp 1987-89 & 2010--18, S \\ 
10560$-$6024 & HDS1561 & 0\farcs3 HIP & 2018, L \\ 
11383$-$6039 & HDS1649 &  0\farcs2 HIP & 2018, L \\ 
12492$-$6040 & HDS1797 &  0\farcs2 HIP & 2018, L \\ 
15037$-$5423 & TDS9389 & 2\farcs2 Tyc & 2018, DR2 \\ 
15066$-$3055 & HDS2128AB & 0\farcs4 HIP & 2016--18, L \\ 
15168$-$1302 & CHR 44 &  0\farcs2 Sp 1983--86 & 2012--18 \\ 
15384$-$1955 & CHR 48 & 0\farcs3 Sp 1983 & 2012--18 \\ 
15470$-$3635 & HDS2223 &  0\farcs13 HIP & 2008--18, L \\ 
15578$-$4100 & SEE 252AB & 0\farcs4 Vis 1897 & 2008--18, L  \\ 
16072$-$2531 & OCC 150 & 0\farcs1 Occ 1931 & 2018, L \\ 
16083$-$2537 & OCC 148 &  0\farcs1 Occ 1931 & 2018, L \\ 
16141$-$1812 & OCC 519 &  0\farcs35 Occ 1977 & 2018, L \\ 
16406$+$0413 & CHR 56Aab & 0\farcs14 Sp 1985--88 & 2008--18, S \\ 
16407$-$6233 & B 1816 & 0\farcs3 Vis 1939 & 2018, L  \\ 
16459$-$3953 & HDS2380 &  0\farcs13 HIP & 2008--18, L  \\ 
17098$-$1031 & TOK 414 & 0\farcs04 Sp 2014 & 2014--18, S \\ 
17146$+$1423 & CHR 139Aab  &   0\farcs2 Sp 1986--91 & 2009--18 \\ 
17341$-$0303 & TOK 417 & 0\farcs1 Sp 2014 & 2015--18, OG  \\ 
18068$+$0853 & TOK 696Aab & 0\farcs03 Sp 2015 & 2015--18, Vib  \\ 
18070$+$3034 & SCA 170Aab &  0\farcs2 2000-05 & 1989--2018, S   \\ 
18112$-$1951 & TOK 57Aab &  0\farcs05 Sp 2008--09 & 2011--18, Vib \\ 
18232$-$2825 & HDS2601 &  0\farcs17 HIP & 2017--18, L \\ 
18267$-$3024 & TOK 421 & 0\farcs07 Sp 2014 & 2014--18, Vib  \\ 
18272$+$0012 & STF2316Aab & 0\farcs2 Sp 1951--2009 & 2008--18, S \\ 
19094$+$1014 & CHR 140 &   0\farcs25 Sp 1985 & 2015--18, L  \\ 
19294$-$0703 & TOK 4Aa,Ab & 0\farcs05 Sp 2009 & 2008--18, Vib \\ 
19488$-$4931 & HDS2818 &  0\farcs17 HIP & 2008--18, L \\ 
19503$+$0754 & CHR 89  & 0\farcs06 Sp 1985--86 & 2017--18, L \\ 
19510$-$0252 & TOK 213Aab & 0\farcs1 Sp 2014 & 2014--18, OG  \\ 
20254$-$2840 & CHR 97 &  0\farcs1  Sp 1983 & 2013--18, S \\ 
20449$+$1219 & B 2910Aab & 0\farcs2 Vis 1937 & 1976--2018, S \\ 
23315$-$2857 & B 602 & 0\farcs2 Vis 1925--32 & 2008--18, L \\ 
23388$-$2816 & B 608 & 0\farcs2 Vis 1925--29 & 2008--18 \\ 
23444$-$7029 & WSI 94 &  0\farcs05 Sp 2008 & 2012--18, Vib \\ 
23598$+$0640 & BAG 31Aab &  0\farcs2 Sp 2001 & 2015--18, S  
\enddata
\tablenotetext{a}{Additional indications of the spurious nature of visual pairs:
DR2 -- parallax provided by {\it Gaia} DR2; 
OG -- optical ghost \citep{SAM17}; 
L -- long estimated period; 
S -- short estimated period or spectroscopic coverage;   
Vib -- artefact caused by telescope vibration. 
}
\end{deluxetable}

Table~\ref{tab:bogus} presents  the list of  candidate spurious double
stars {\bf observed at SOAR.}  Its  first two columns link the pair to
the WDS  catalog \citep{WDS}.  Column (3) describes  the resolution by
giving  the separation  in arcseconds,  measurement technique  (Vis --
visual, Sp -- speckle, HIP -- {\it Hipparcos}, Tyc -- {\it Tycho}, Occ
-- lunar occultations), and the years when the pair was resolved.  The
last  column gives  the years  of  non-resolutions {\bf  at SOAR}  and
additional hints  coded by  letters.  DR2 indicates  non-resolution by
{\it Gaia}  (resolved binaries do  not have parallaxes in  DR2).  Many
objects  are located  at large  distances, and  their  separations, if
real, imply periods  of $>$100 years (code L).   Similarly, the code S
means that  the period  of non-resolution is  comparable to  the short
estimated  binary  period;  in  some  cases  the  spectroscopic  orbit
provides a strong evidence against existence of close visual binaries.
False  resolutions at  SOAR are  explained, mostly,  by the  effect of
vibration {\bf  \citep[see Figure~6 of][]{HRCAM} } that  was not fully
appreciated during the  first years of HRCam operation  (code Vib) and
by  optical  ghosts  (code   OG).   {\bf  Although  some  observations
  presented  here  are  still  affected  by  vibration,  this  is  now
  recognized and  compensated for by  the use of reference  stars with
  similar  artefacts.  Several  objects  in Table~\ref{tab:bogus}  are
  unresolved   subsystems  in   visual  triple   stars,   while  their
  successfully measured pairs are found in Table~\ref{tab:measures}. }

\begin{figure}
\epsscale{1.1}
\plotone{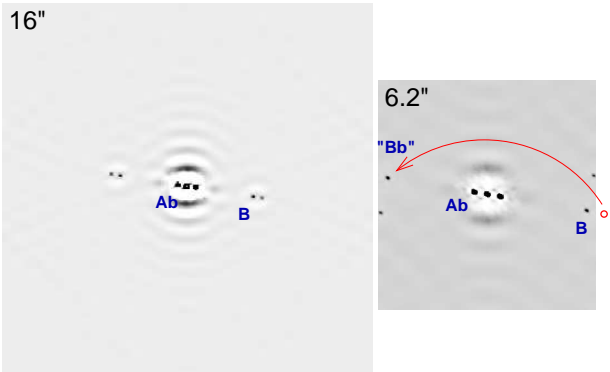}
\caption{Triple system  07185$-$5724. The left panel shows  ACF of the
  data  cube  recorded  in  2018.2  with the  full  field;  the  peaks
  corresponding to the companions Ab at 0\farcs4 and B at 2\farcs9 are
  indicated.  In the right  panel, the  ACF in  the 6\farcs2  field is
  shown. The  peak corresponding to  the correlation between Ab  and B
  (red circle) falls outside the field and is aliased (reflected) to a
  different  position (red  arrow),  creating an  illusion of  another
  companion Bb.
\label{fig:RST244} }
\end{figure}

The   WDS   J07185$-$5724   (RST~244   Ba,Bb)  is   a   special   case
(Figure~\ref{fig:RST244}).   The  pair  Ba,Bb  in a  visual  quadruple
system  HIP~35374 (Aa,Ab  is a  0\farcs4 binary)  was  discovered with
HRcam  at SOAR  at 0\farcs9  separation in  2010 and  measured several
times since.   It was observed here  with the full field  to show that
the  companion Bb  corresponds  to the  peak  produced by  correlation
between B and Ab; it is aliased and appears at the wrong position when
the data cubes with the 6\farcs2 field are recorded. Consequently, the
pair Ba,Bb does not exist, and this  system is only a triple.


\section{Summary}
\label{sec:sum}

Continued  monitoring  of  close  visual  binaries  at  SOAR  makes  a
substantial contribution to the definition of their orbits, especially
for  tight   and  nearby  pairs  with  short   periods  like  HIP~6626
(Figure~\ref{fig:HIP6626}).   Good-quality  visual  orbits coupled  to
precise parallaxes from {\it Gaia} will vastly extend our knowledge of
stellar  masses.  Moreover, visual  orbits  are  needed in  different
astrophysical  contexts.   To  give  an  example,  our  orbit  of  the
exo-planet  host HIP~49522  (10067+1754)  with $P=203$  years is  still
poorly  constrained,  but  the  premature  51-year  orbit  proposed  by
\citet{MaB2016} is certainly refuted by our measurements, resolving the apparent
conflict with the planetary orbits discussed in the above paper.

The SOAR speckle  program resulted in the discovery  of many new close
binaries  and subsystems.  This  list is  extended here  by the  35 new
subsystems in visual multiples,  newly resolved {\it Hipparcos} stars,
and tight PMS binaries in  Orion OB1. SOAR speckle observations of PMS
stars  in  various   nearby  star-forming regions  are  a  key   part  of  our
multiplicity  studies  in   young  stellar  populations,  probing  the
separation  regime $\sim$30--1000 AU  at $\sim$400\,pc.   Understanding the
formation of stellar systems requires comprehensive multiplicity census
of PMS stars across regions with differing conditions.

During  2018,   the  core  program   on  visual  multiples   has  been
supplemented by  various binary surveys;  high-resolution screening of
TESS exo-planet candidates has started as well.  These programs will be
continued  and their  results  will be  published  in   forthcoming
papers.


\acknowledgments 

We thank the SOAR operators for efficient support of this program, and
the SOAR director J.~Elias for  allocating some technical time. {\bf A
  through checking by the anonymous  Referee has helped to improve the
  presentation and to correct some errors.}

This work  is based  in part on  observations carried out  under CNTAC
program CN2018A-2. R.A.M. acknowledges support from the Chilean Centro
de Excelencia en Astrof\'{i}sica y Tecnolog\'{i}as Afines (CATA) BASAL
AFB-170002, and FONDECYT/CONICYT grant \# 1190038.

This work  used the  SIMBAD service operated  by Centre  des Donn\'ees
Stellaires  (Strasbourg, France),  bibliographic  references from  the
Astrophysics Data  System maintained  by SAO/NASA, and  the Washington
Double Star  Catalog maintained  at USNO.  This  work has made  use of
data   from   the   European   Space   Agency   (ESA)   mission   Gaia
(\url{https://www.cosmos.esa.int/gaia}  processed  by  the  Gaia  Data
Processing      and     Analysis      Consortium      (DPAC,     {\url
  https://www.cosmos.esa.int/web/gaia/dpac/consortium} Funding for the
DPAC  has been provided  by national  institutions, in  particular the
institutions participating in the Gaia Multilateral Agreement.

{\it Facilities:}  \facility{SOAR}.



\end{document}